\documentclass{article}

\usepackage{arxiv}

\usepackage{caption}

\usepackage{multirow}

\newcommand{\ignore}[1]{}

\usepackage{charter}
\usepackage{pifont}
\usepackage{amsmath,amsfonts,amssymb,amsthm}
\usepackage{epsfig}
\usepackage{amsmath}
\usepackage{setspace}
\usepackage{graphicx}
\usepackage[toc,page]{appendix}

\usepackage{subfigure}
\usepackage[numbers, sort&compress]{natbib}
\usepackage[utf8]{inputenc}
\usepackage[english]{babel}
\usepackage{amsmath,amssymb,amsfonts}
\usepackage{bm}
\usepackage{textcomp}
\usepackage[T1]{fontenc}
\usepackage{graphicx}
\usepackage{float}
\usepackage{geometry}
\usepackage{fullpage}
\usepackage[Conny]{fncychap}
\usepackage{amsmath}
\usepackage{color}
\usepackage{setspace}
\usepackage{graphicx}
\usepackage{qtree}
\usepackage{tikz}
\usepackage{pgf}
\usepackage{enumitem}
\usepackage{empheq}
\usepackage{url}
\usepackage[colorinlistoftodos]{todonotes}

\newcommand{\xmark}{\ding{55}}
\newcommand{\ket}[1]{|{#1}\rangle}

\graphicspath{ {./images/} }

\title{General Classification of Entanglement Using Machine Learning}

\author{
 F. El Ayachi and M. El Baz \\
  ESMaR,Faculty of Sciences\\
  Mohammed V University in Rabat\\
  Av. Ibn Battouta, B.P. 1014, Agdal, Rabat, Morocco. \\
  \texttt{fadwa\_elayachi@um5.ac.ma} \\
  \texttt{morad.elbaz@um5.ac.ma}
  }

\begin{document}
\maketitle
\begin{abstract}
A classification of multipartite entanglement in qubit systems is introduced for pure and mixed states. The classification is based on the robustness of the said entanglement against partial trace operation. Then we use current machine learning and deep learning techniques to automatically classify a random state of two, three and four qubits without the need to compute the amount of the different types of entanglement in each run; rather this is done only in the learning process. The technique shows high, near perfect, accuracy in the case of pure states. As expected, this accuracy drops, more or less, when dealing with mixed states and when increasing the number of parties involved.
\end{abstract}

% keywords can be removed

%%%%%%%%%%
\section{Introduction}
\label{sec:intro}

Entanglement \cite{EPR,Entshro,onEPR}, as a quantum resource, simplifies and in many cases simply make it possible, the execution of some quantum information tasks. It has many applications, in quantum computing \cite{Nielsen}, quantum communications \cite{telepo93, densecoding92} , quantum cryptography \cite{QKDMult}, security devices with quantum abilities \cite{radargerry, QRadar}... 

Many advances are made in the direction of understanding this purely quantum feature \cite{Revhored2009}. But in spite of these, only the bipartite entanglement in two dimensional systems, {\textit{i.e.}} qubits, has a satisfactory description and a proper one is still lacking when moving to higher dimensional subsystems, or when increasing the number of qubits composing the overall system. In the latter case, when studying the entanglement, the classification of entangled states is much richer than in bipartite systems. Instead of the simple distinction of being separable or entangled for bipartite states, one has to specify for the latter whether they are partially entangled or fully entangled states when dealing with multipartite systems. Moreover, the number of combinations of the different components increases and consequently the number of ways of sharing the entanglement among the components exponentially increases. Each way of achieving this then might define a different class of multipartite entanglement \cite{class3qubits, Distent}; although that is not the only way of defining a class of multipartite entanglement. As a matter of fact, there exists several classifications that are based on other properties of multipartite entanglement, such as the classification based on local operations and classical communications (LOCC) \cite{class3qubits}, it's extension to stochastic LOCC (SLOCC) \cite{class4qubits}, or classification based on persistent homology \cite{multimancini}. 

Achieving a well defined classification of multipartite entanglement is of paramount importance, as it is established that quantum information processing tasks based on multipartite entanglement are sensitive to the kind or class of entanglement being used (\cite{QKDMult}, \cite{QRadar}, \cite{qsistate}). In this regard, each task actually requires a specific feature out of the multipartite entanglement on which it is based, such as how it behaves when one or more components are out of play (traced out). As such, it is a prerequisite to know the class to which a given state belongs in order to assert its relevance to a given quantum task.

In recent years, several studies focused on combining the disciplines of quantum information processing \cite{Nielsen} and machine learning \cite{Mitchell15ML} resulting in the new discipline of quantum machine learning \cite{peter, superQMLshuld, QML}. This combination can be achieved in one of four possible ways. One can develop quantum versions of machine learning algorithms and use them to perform either classical tasks or quantum ones \cite{QMLShuld2014}. On the other hand, more recently, it was demonstrated that simply using classical machine learning algorithms can be useful in solving quantum tasks especially classification tasks \cite{NNclasscore,steerdetec}; this requires using classical algorithms in order to manipulate quantum data. This is the approach adopted in the current work for which we chose deep learning neural networks \cite{NielsenML,NN} and evaluate its performance in the classification of full multipartite entanglement of qubits in both pure and mixed states.

To achieve this, we start by presenting in section \ref{sect.calss}  the classification of fully entangled states adopted in this paper and how it naturally emerges from the use of tangle as a measure of multipartite entanglement. The classification tree resulting and how it translates to the particular case of three and four qubits as well as a discussion of this classification scheme are also given. Section \ref{sect.MLcalss}, contains the strategy adopted in using deep learning artificial neural networks to implement the classification of fully entangled multipartite states. We present the dataset used for the learning and testing phase of the protocol. A presentation of the results in terms of the confusion matrix and table of verification is given at the end of the section, followed by a conclusion with some perspectives of the current work. Appendices are added at the end, where the definitions of the entanglement measures used are given as well as the definitions of the multipartite entangled coherent states and their properties. Particular cases of each of the different classes of fully entangled states for three and four qubits are presented in terms of these coherent states. Their dependence on the amplitude parameter, simplifies the presentation and discussion of the relevant properties of the different classes discussed in the main text.

\section{Classification of fully-entangled states}\label{sect.calss}
An arbitrary quantum system $A_{1}\otimes...\otimes A_{N}$, with $N$  subsystems, described by the density matrix $\rho \in H = H^{A_1} \otimes ... \otimes H^{A_N} $  is called fully-separable if and only if $\rho$ can be written in the form $\rho = \rho_{A_1} \otimes ... \otimes  \rho_{A_N} $, where $\rho_{ A_i}$ are reduced density matrices of the individual subsystems from $H^{A_i} $. On the other hand if $\rho$ can be written as a tensor product of $k$ substates: $\rho = \rho_{1} \otimes ... \otimes  \rho_{k}   $ with $k < N$ then it is called $k$-separable \textit{i.e.}  some subsystems are separable from the rest that is entangled. Outside of these two cases  $\rho$ is called fully-entangled, in which case all parties of the system are entangled.

In this paper, we focus our attention on fully-entangled multipartite systems as the discussion of $k$ separable states classification is always amenable to that of fully entangled ones in a lower dimensional Hilbert space.

\subsection{Entanglement quantifiers and detectors }
\label{Entquant}
The entanglement in a representative state, of a given class of entanglement (defined later in subsection \ref{sbsect_tree}), is evaluated using the $i$-tangles, $\tau_{i}$. In the following we present their definitions in a general form (valid for both pure and mixed states) adapted for the task at hand:

	\begin{eqnarray}
	\tau_{1} &=& \left(\prod_{p} \sum_{\alpha\beta }(C^{p}_{\alpha, \beta} )^{2} \right) ^{\dfrac{1}{m}}
    \label{to1},\\
	\tau_{2} &=& \dfrac{1}{\binom{N}{2}} \sum_{i, j } (C_{i j})^{2}\label{to2},\\
	\tau_{3} &=&  \dfrac{1}{\binom{N}{3}}\sum_{i, j,k }\left( \dfrac{1}{3}  \sum_{\alpha, \beta }  [\,(C^{ij/k}_{\alpha\beta})^{2} + (C^{ik/j}_{\alpha\beta})^{2} + (C^{jk/i}_{\alpha\beta})^{2} ]\right) \label{to3},\,  \\
	\tau_{4} &=& \dfrac{1}{\binom{N}{4}} \Bigg(\sum_{i, j,k,l } (\dfrac{1}{7}  \sum_{\alpha, \beta } [\,(C^{i/jkl}_{\alpha\beta})^{2} + (C^{j/ikl}_{\alpha\beta})^{2} + (C^{k/jil}_{\alpha\beta})^{2} +(C^{ijk/l}_{\alpha\beta})^{2} \nonumber \\  & & +(C^{ij/kl}_{\alpha\beta})^{2} + (C^{ik/jl}_{\alpha\beta})^{2} +(C^{il/kj}_{\alpha\beta})^{2}]) \label{to4} \, \Bigg), \\
	&\vdots&\nonumber \\
	\tau_{N} &=& \dfrac{1}{\binom{N}{N}}  \left(\dfrac{1}{m} \sum_{p } \sum_{\alpha,\beta }  [\,(C^{p}_{\alpha\beta})^{2} ]\right) \label{toN},\, 
	\end{eqnarray}
    with $m = 2^{N-1} -1$, $N$ being the number of qubits and $\binom{N}{i}$ the binomial coefficient, giving all possible $i$-tuples out of $N$ qubits. 
    
    For the sake of not overloading the main text with additional definitions, we put  those of the different concurrences used above as well as the corresponding references in the appendix \ref{appendix1}. For instance, $C_{i/j}$ is the Wooter's concurrence (\ref{Wconc}) representing 2-qubit bipartite entanglement, while $C^{p}_{\alpha\beta}$ represents the  entanglement shared in a bipartition $p$, with an arbitrary dimension for each partition. It is given by the I-concurrence (\ref{iconc}) when $\rho_N$ is a pure state;  if it is a mixed state $C^{p}_{\alpha\beta}$ is based on the lower bound (\ref{clb}). 
	
	As such the $i$-tangle is defined as weighted sum of the squares of the concurrences in each possible bipartition of the $i$ parties.
	
	The product $\prod_{p}$  over all possible combinations, is adopted in equation (\ref{to1}) as it is more suitable to verify if a given state is fully entangled or not using $\tau_{1}$.
	
	In the case of pure states (of three and four qubits), it is better to use the \textit{pure-states-tangle} $\tau_{3}$ (\ref{t3pure}) and $\tau_{4}$ (\ref{t4pure}) quantifying respectively the 3-way and 4-way entanglement, because of their ability to quantify  more precisely genuine entanglement.
	
	For $\tau_{N}$, the $\sum_{p}$ is carried over all possible partitions $p$; this ensures that for fully separable multipartite states, $\tau_{N} = 0$. Thus allowing $\tau_{N} > 0$ to indicate the presence of at least some type of entanglement inside the quantum state.

\subsection{Different types of entanglement}
Our classification scheme is based on how "fragile" the fully entangled states are when we apply a partial trace operation on them. To quantify this entanglement we use the $i$-tangles defined in the previous subsection \ref{Entquant}.

Let $\rho \in H = \underset{\times N}{\underbrace{H^{2} \otimes ... \otimes H^{2}}}  $  be a (pure or mixed) state of $N$ parties; then $\rho$ can be fully-entangled in $N-1$ nonequivalent ways:
\begin{itemize}
	\item the $N$-entanglement, in which case performing a partial trace over any individual subsystem will yield a separable state. Such an entanglement is shared between all the $N$ parties.
	\item The $(N-1)$-entanglement, which vanishes if we trace out any two individual subsystems. This entanglement is shared between $(N-1)$-tuples.
	$$\vdots$$
	\item The $2$-entanglement, is the most robust against partial trace as it does not vanish even if traced out over any $(N-2)$ subsystem. In this case the entanglement is shared between all possible pairs.
\end{itemize}

More generally  the $(N-i)$-entanglement, vanishes  if we trace over any $(i+1)$ qubits and the entanglement is shared between $(N-i)$-tuples. Henceforth, this entanglement is quantified using the $(N-i)$-tangle ($\tau_{N-i}$) as defined in the previous subsection.

It is important to note that a fully entangled state can contain one type of entanglement or more, which imposes defining an order of precedence in this classification. This is carried over in the next subsection.

\subsection{Classification tree}
\label{sbsect_tree}
A preliminary classification of an arbitrary $N$-qubit state, puts it in a category of fully separable, $k$-separable or fully entangled states. Then, if it is fully entangled, one can check the type of entanglement in play, by first checking if it contains the 2-entanglement as defined in the previous subsection. If it is the case, it is placed in the $[N]_{2}$ category, if not one checks whether it contains 3-entanglement, in which case it is a $[N]_{3}$ entanglement class and so on up to the last one where only states containing only the $N$-entanglement type as defined in the previous subsection are left, which form the $[N]_{N}$ class of fully entangled states. This last class, should be the most fragile under partial trace operations.

\begin{figure}[h]
\begin{center}
\begin{tikzpicture}[sibling distance=10em,
	every node/.style = {shape=rectangle, rounded corners,
		draw, align=center,
		top color=white, bottom color=pink!60}]]
	\node {N-qubits density matrix}
	child { node {$ \tau_{1} = 0 $}
	    child{node{separable \\or\\ $k$-separable} } }
    child { node {$ \tau_{1} \neq 0 $}
	    child{node{Fully entangled} 
	        child {node {$ \tau_{2} \neq 0 $}
	            child{node{$[N]_{2}$} }}
	        child {node {$ \tau_{2} = 0 $}
	            child{node{$ \tau_{3} \neq 0 $}
	                child{node{$[N]_{3}$} }}
	            child{node{$ \tau_{3} = 0 $}
	                child{node{$ \tau_{4} \neq 0 $}
	                    child{node{$[N]_{4}$} }}
	                child{node{$ \tau_{4} = 0 $}
	                    child{node{$\vdots$}
	                        child{node{$ \tau_{N} \neq 0 $} 
	                            child{node{$[N]_{N}$}}}
	                            }
	                            }
	                            }
	                }}
	       };
\end{tikzpicture}
\caption{Classification of entanglement for $N$-qubit states}
\label{fig:tree}
\end{center}
\end{figure}
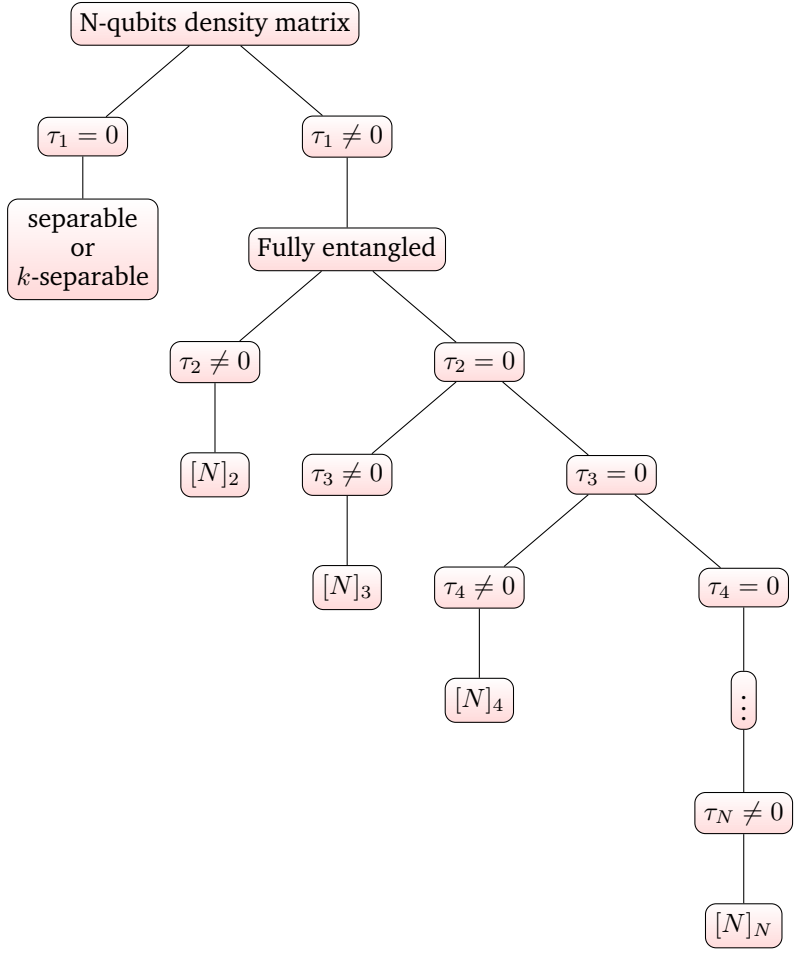

This procedure is schematized using a hierarchical tree showed in Figure \ref{fig:tree}.

\subsection{Classification of pure states }
\label{sbsect_class}
\subsubsection{Three qubits }
Applying the previous classification of entanglements to the case of three qubits, yields two classes of fully entangled states that are represented by the well-known $W$ and $GHZ$ sates. This classification is compatible with a LOCC\footnote{Local operation and classical communication}-based classification in \cite{class3qubits}. In the following we will represent the different representative states ($W$ and $GHZ$ sates and others when dealing with systems containing more than three qubits) in the coherent state basis $\left\{ \ket{\alpha} \right\}$; this will have the advantage of allowing to plot the different relevant $i$-tangles in terms of the amplitude $\alpha$, keeping in mind that we recover the standard ($W$ and $GHZ$) sates written in the computational basis when $\alpha$ (or equivalently the mean photon number) is large enough. The definitions of coherent states used are given in Apprendix \ref{appendix2}. 

\begin{figure}[h]
    \centering
    \subfigure[]{\includegraphics[width=0.3\linewidth]{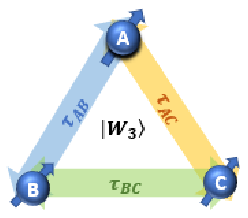}\label{threeW}} 
    \subfigure[]{\includegraphics[width=0.4\linewidth]{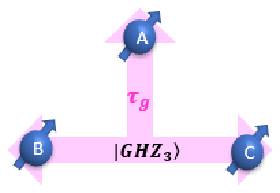}\label{threeGHZ}} 
    \subfigure[]{\includegraphics[width=0.4\linewidth]{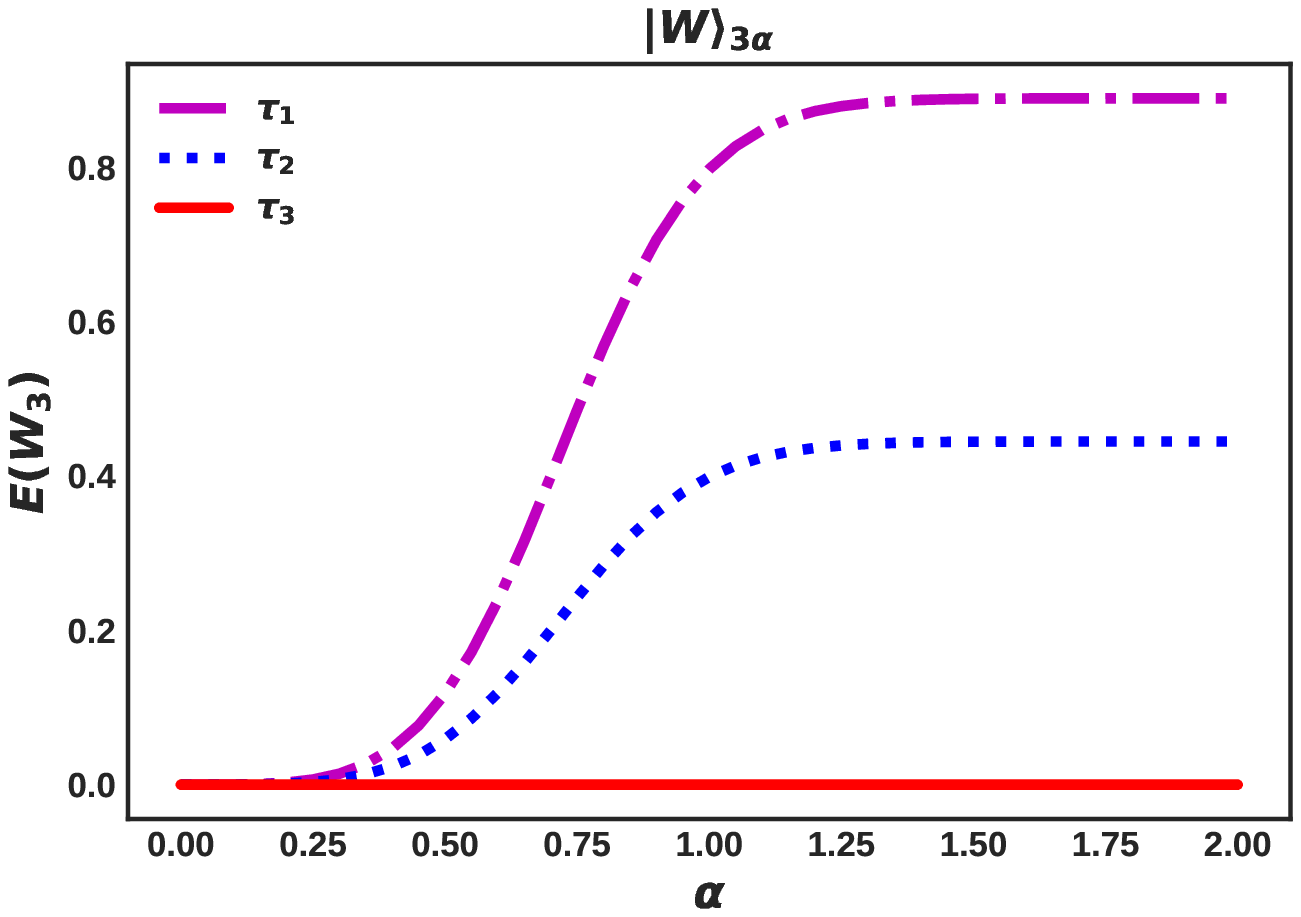}\label{3entW}}
    \subfigure[]{\includegraphics[width=0.4\linewidth]{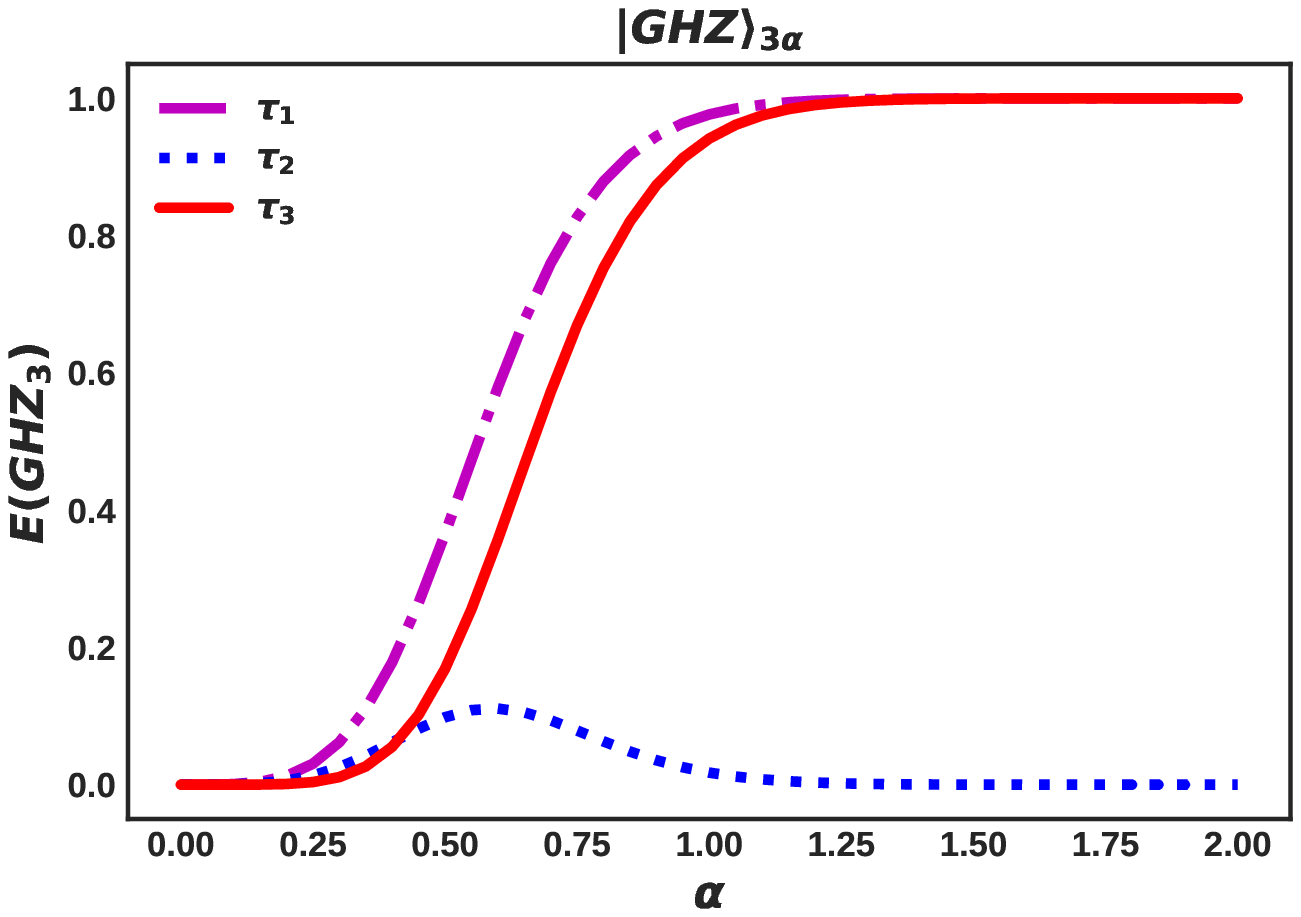}\label{3entGHZ}}
\caption{Entanglement of the representative states of the classes of  fully-entangled three-qubit states. (a) and (b) are respectively descriptive schemes corresponding to  the representative states of the classes $[3]_{2}$ and $[3]_{3}$. (c) and (d) show the amount of each type of entanglement present in each of the representative states.}
    \label{fig:three}
\end{figure}

Figure \ref{fig:three},  shows the different types of entanglement that are present in the two classes represented by the $W$ and $GHZ$ sates as well as the amount of each type of entanglement involved.

The first class $[3]_{2}$, schematized in Figure \ref{threeW}, contains the states for which the entanglement is shared between each pair ($i,j$) \textit{i.e.} states with a non-zero two-tangle, $\tau_{2}$ as defined in equation (\ref{to2}). The representative state for this class is the $W$-state defined by
$\arrowvert W_{3} \rangle = \dfrac{1}{\sqrt{3}}( \arrowvert 100 \rangle + \arrowvert 010 \rangle + \arrowvert 001 \rangle)$. Figure \ref{3entW} based on the coherent state version of the $W$-states (see equation (\ref{W3alpha}) in Appendix \ref{appendix2}), shows that in this states the global entanglement $(\tau_{1})$ as defined in equation (\ref{to1}) is double the value of the bipartite one ($\tau_{2}$). This is due to the fact that the former quantifies the entanglement based on a (qubit, two-qubits) bipartition of the system while the later takes into account all (qubit,qubit) bipartitions. For instance, the global entanglement based on a ($A,BC$) bipartition takes into account the entanglement between the two pairs ($A,B$) and ($A,C$) and no genuine tripartite entanglement is present. This type of  entanglement holds up even after a measurement is performed over one of the three qubits.

The second class $[3]_{3}$ is represented in Figure \ref{threeGHZ}, where the entanglement is shared between all parties. This class contains all states with zero two-tangle ($\tau_{2}$) and non zero three-tangle ($\tau_{3}$) and can be represented by the GHZ-state defined as
$\arrowvert GHZ_{3} \rangle = \dfrac{1}{\sqrt{2}}( \arrowvert 000 \rangle + \arrowvert 111 \rangle)$. Figure \ref{3entGHZ} based on the coherent state version of the $GHZ$-states (see equation (\ref{GHZ3alpha}) in Appendix \ref{appendix2}), shows that in this state the global entanglement $(\tau_{1})$ equals the genuine tripartite entanglement ($\tau_{3}$) and no bipartite entanglement is present. This type of entanglement vanishes if a measurement is performed over one of the qubits.

\subsubsection{Four qubits}
It is well established that as the number of qubits increases, the classes of entanglement increase as well. However these classes might differ from one work to another depending on the defining treat adopted for distinguishing the classes. Based on our classification, the case of four qubits yields three classes of fully entangled states, two of which are the same for the three qubits case. Namely, in the four qubit case, the classes obtained are represented by the $W$ and $GHZ$ versions for four qubits, in addition to another class we will designate as the $X$-state.

\begin{figure}[h]
    \centering
    \subfigure[]{\includegraphics[width=0.35\linewidth]{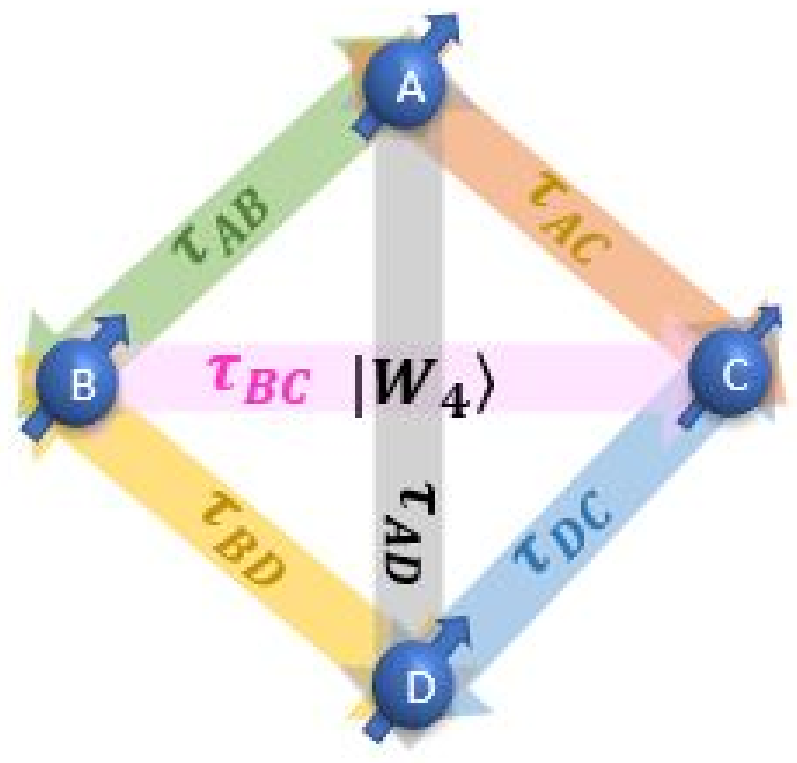}\label{FourW}} 
    \subfigure[]{\includegraphics[width=0.29\linewidth]{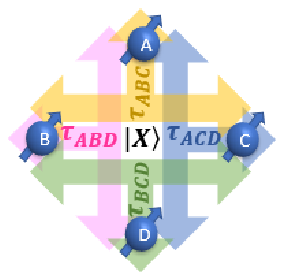}\label{FourX}} 
    \subfigure[]{\includegraphics[width=0.33\linewidth]{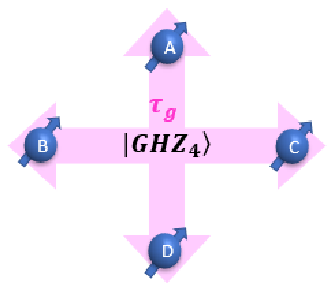}\label{FourGHZ}} 
    \subfigure[]{\includegraphics[width=0.325\linewidth]{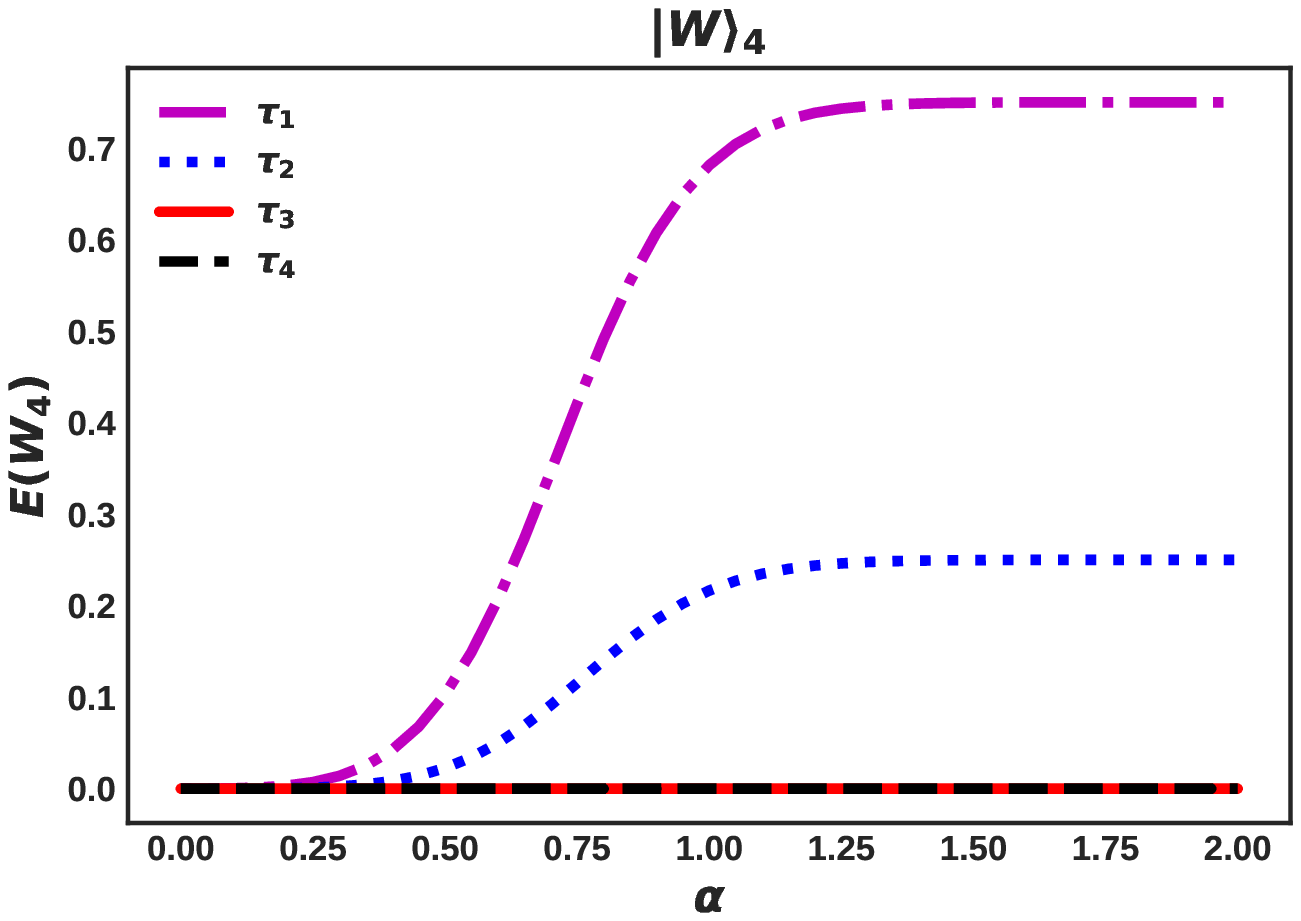}\label{4entW}}
    \subfigure[]{\includegraphics[width=0.325\linewidth]{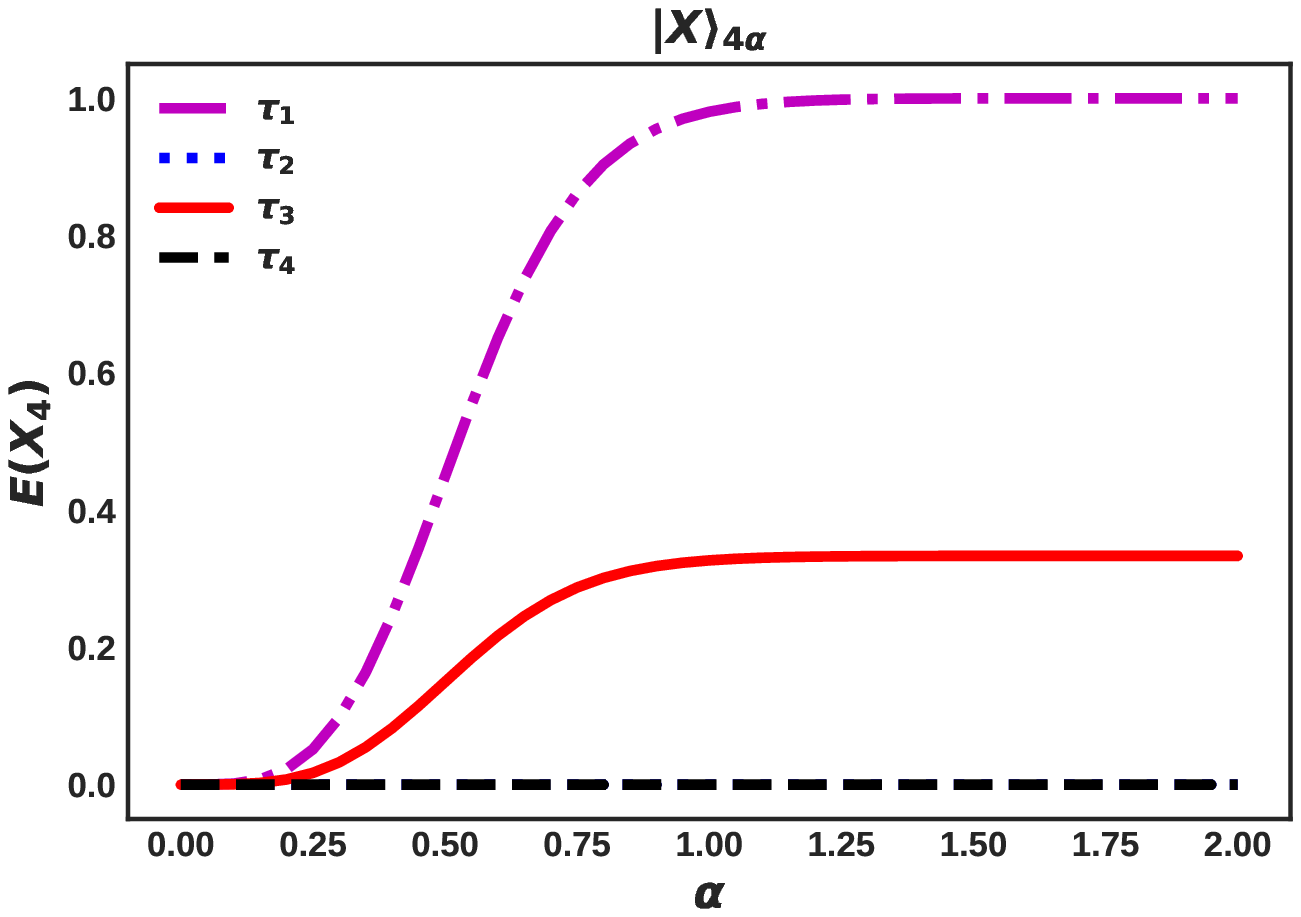}\label{4entX}}
    \subfigure[]{\includegraphics[width=0.325\linewidth]{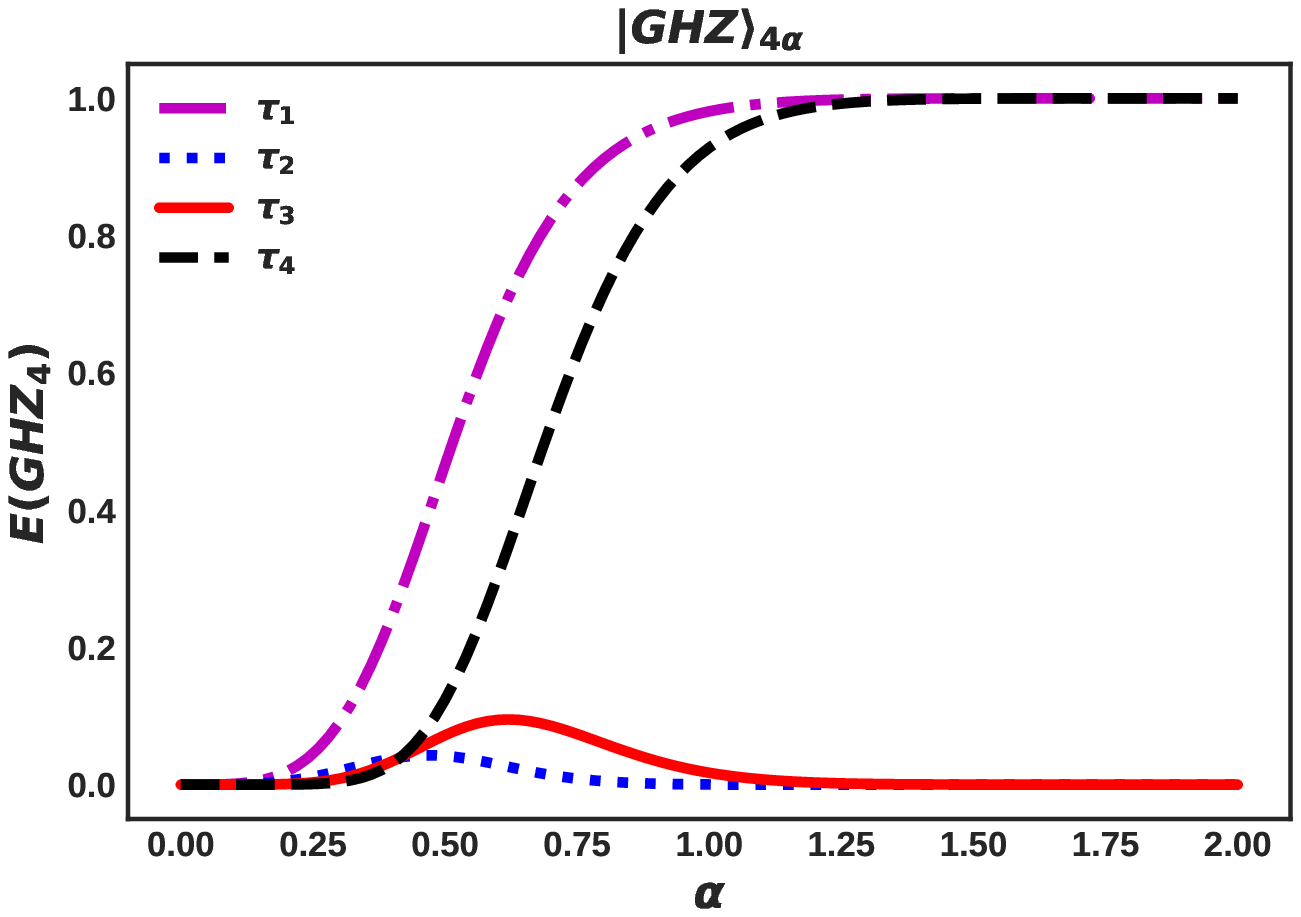}\label{4entGHZ}}
\caption{Entanglement of the representative states of the classes of  fully-entangled four-qubits states. (a), (b) and(c) are respectively descriptive schemes corresponding to  the representative states of the classes $[4]_{2}$, $[4]_{3}$ and $[4]_{4}$  . (d), (e) and (f) show the amount of each type of entanglement present in each of the representative states.}
    \label{fig:Four}
\end{figure}

Figure \ref{FourW} represents the $[4]_{2}$ category, that contains all states with non-zero two-tangle ($\tau_{2}$) between any two pairs $(i,j)$. The representative state of this class is the $W$-state version for four qubits, defined by $ \arrowvert W_{4} \rangle = \dfrac{1}{2}( \arrowvert 1000 \rangle + \arrowvert 0100 \rangle + \arrowvert 0010 \rangle + \arrowvert 0001 \rangle) $. Figure  \ref{4entW} based on the coherent state version of this later (see equation (\ref{W4alpha}) in Appendix \ref{appendix2}),  shows that in this state only the ($\tau_{2}$) tangle is present and it combines to the global entanglement of four qubits $(\tau_{1})$  (\ref{to1}) the value of which is triple  that of the bipartite one ($\tau_{2}$) as it is shared between three pairs $(A,B)$, $(A,C)$ and $(A,D)$, so this type of  entanglement  hold up even when tracing out one or two of the four qubits.

The second Figure \ref{FourX} represents the $[4]_{3}$ class that contains all states with a zero two-tangle ($\tau_{2}$) between each pair $(i,j)$ and non-zero three-tangle ($\tau_{3}$)  between all triplets $(i,j,k)$. The representative state for this class is the state $ \arrowvert X \rangle = \dfrac{1}{\sqrt{5}} ( \arrowvert 0000 \rangle + \arrowvert 0111 \rangle + \arrowvert 1011 \rangle +\arrowvert 1101 \rangle + \arrowvert 1110 \rangle) $. Figure \ref{4entX} based on the coherent state version of this latter (see equation (\ref{X4alpha}) in appendix \ref{appendix2}), shows that for this state only ($\tau_{3}$) is present while the other types of entanglement are absent. This results in this type of entanglement surviving when one measures one of the four qubits but vanishes if the measurement is performed on a second qubit.

The last Figure \ref{FourGHZ} represents the class $[4]_{4}$  in which the entanglement is shared between all components. This class contains all states with only a four-tangle ($\tau_{4}$) ( \ref{t4pure}) while the two and three tangle ($\tau_{2},\tau_{3}$) are zero. The representative state for this class is the $GHZ$ version for four qubits, which is defined by $ \arrowvert GHZ_{4} \rangle = \dfrac{1}{\sqrt{2}}( \arrowvert 0000 \rangle + \arrowvert 1111 \rangle$). Figure \ref{4entGHZ}, based on the coherent state version of this later (see equation (\ref{GHZ4alpha})), shows that in this state the global entanglement $(\tau_{1})$ equals the ($\tau_{4}$), so this type of entanglement vanishes if we trace out any of the four qubits.

\subsection{Classification of mixed states }
To study the types of entanglement that are contained in a given random mixed state, we start by analyzing the mixture of  the representative states defined in the previous subsection and plot their $i$-tangles defined in equations (\ref{to2},\ref{to3} and \ref{to4}). This is applied for three and four qubits cases below.

\subsubsection{Three qubits}
 For three qubits, we write a general mixture of $\arrowvert GHZ_{3} \rangle$ and $\arrowvert W_{3} \rangle$ states as
\begin{equation}
 \rho(b) = b \arrowvert GHZ_{3} \rangle \langle GHZ_{3} \arrowvert + (1-b) \arrowvert W_{3} \rangle \langle W_{3} \arrowvert,
    \label{mghzwstate}
\end{equation}
with $b$ being the mixing parameter ranging from $0$ to $1$.

\begin{figure}[h]
	\centering
		\includegraphics[width=0.6\linewidth]{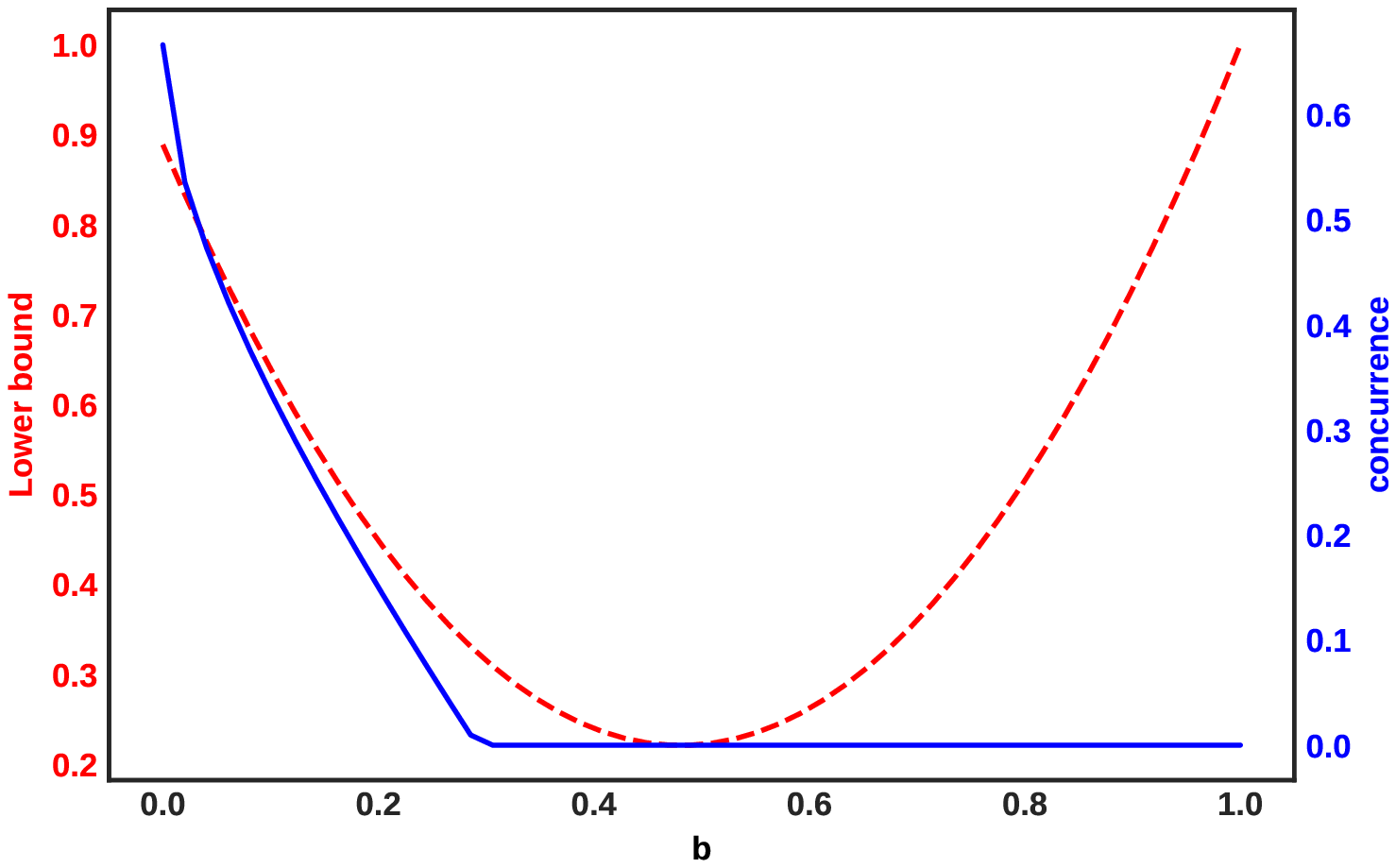}
	\caption{ Entanglement with respect to the mixing parameter $b$ for mixtures of representatives states in three qubits, the solid blue line represents the bipartite concurrence and the dashed red line the lower bound for tripartite entanglement.}
	\label{mixedghzw3}
\end{figure}

It is clear that the type of entanglement, present in the mixed state (\ref{mghzwstate}) of $GHZ$ and $W$, depends on the mixing parameter $b$ such that, when $b\le 0.3$, figure \ref{mixedghzw3} shows that the state belongs to $[3]_{2}$ class, because the bipartite concurrence is non-null. On the other hand when $b>0.3$, figure \ref{mixedghzw3} shows that the concurrence is absent but the lower bound is still present which places the state in the $[3]_{3}$ class.
\begin{figure}[h]
    \centering
    \subfigure[$b=0.2$]{\includegraphics[width=0.3\linewidth]{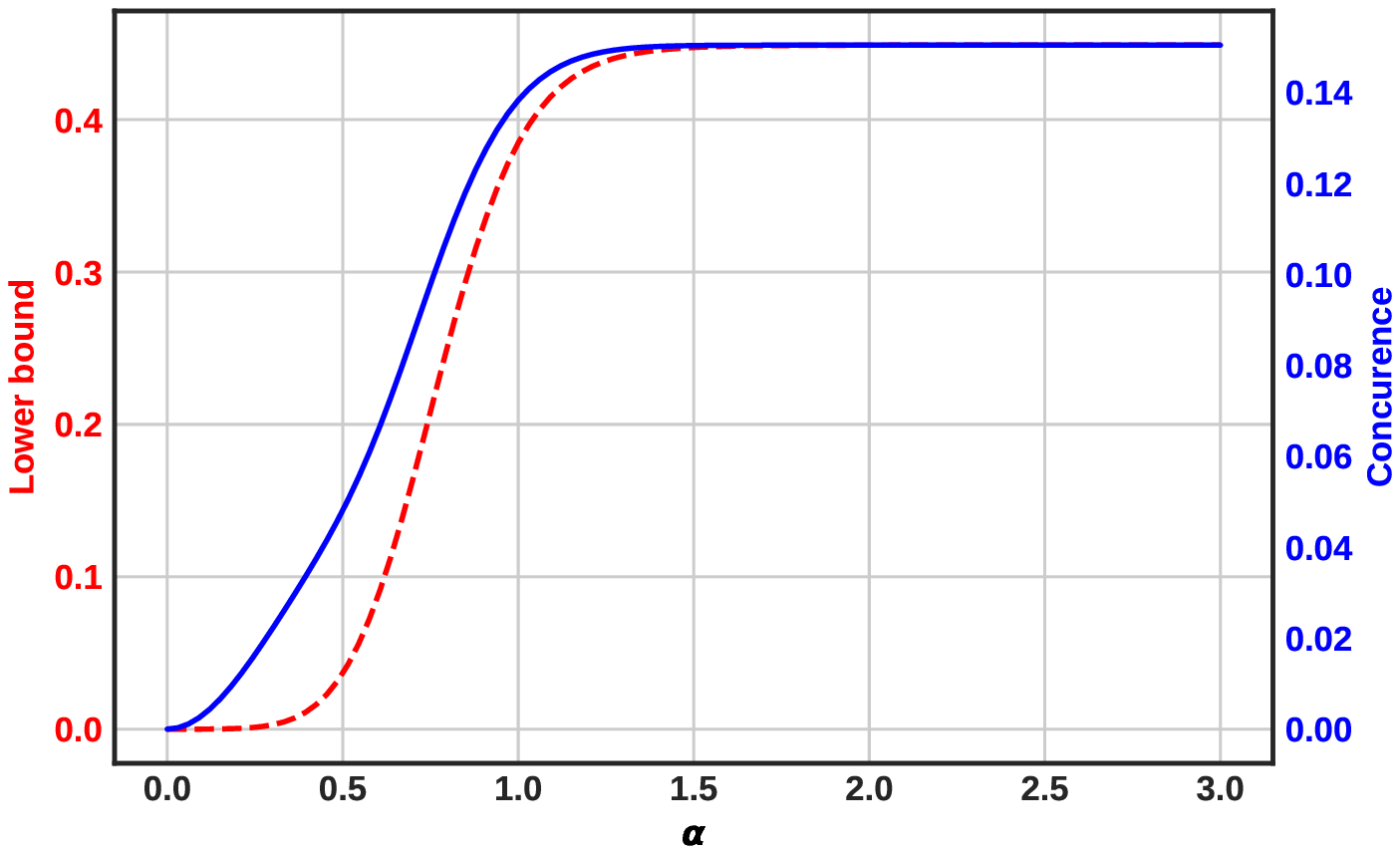}\label{ghzw02}}
    \subfigure[$b=0.5$]{\includegraphics[width=0.3\linewidth]{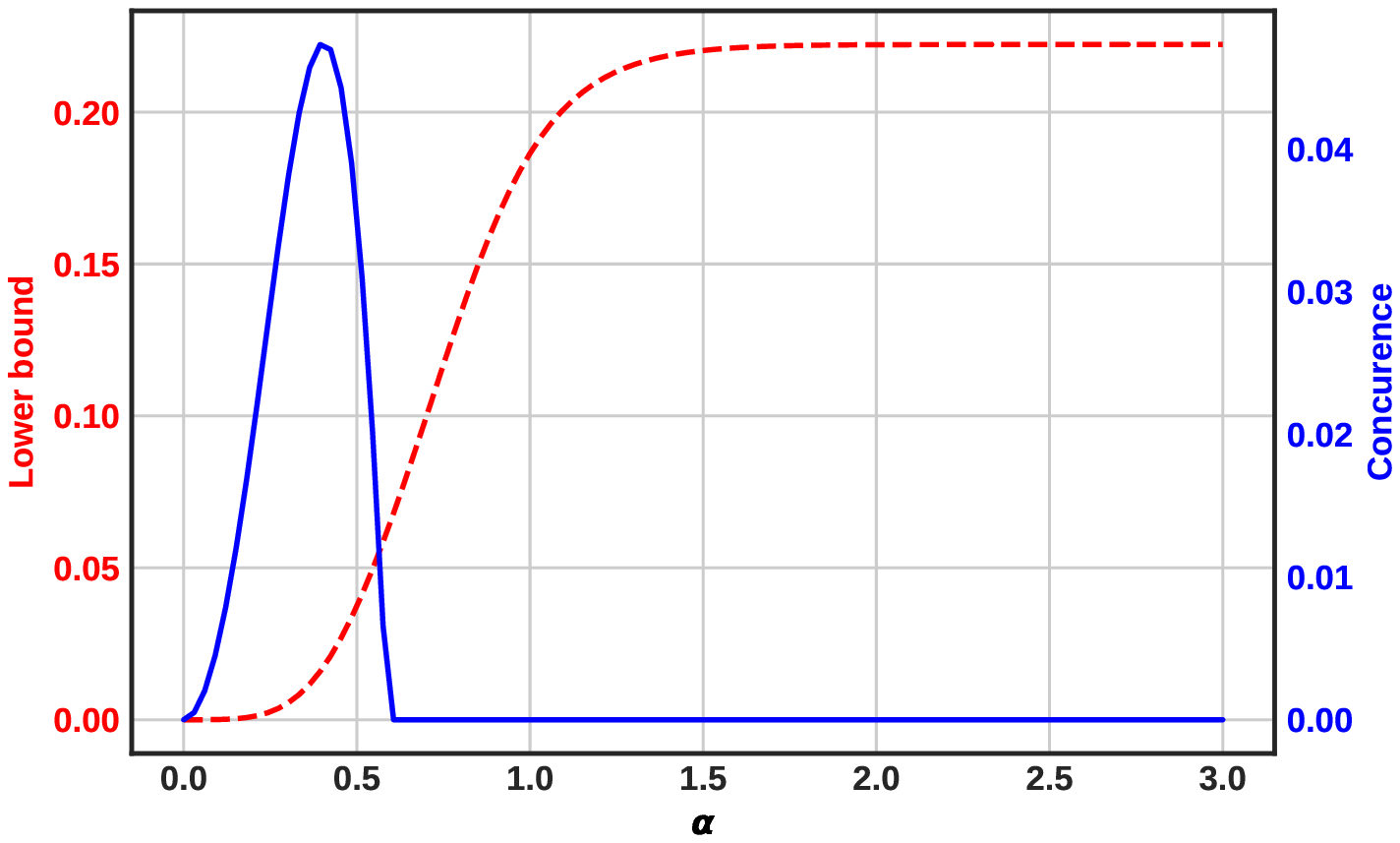}\label{ghzw05}}
    \subfigure[$b=0.8$]{\includegraphics[width=0.3\linewidth]{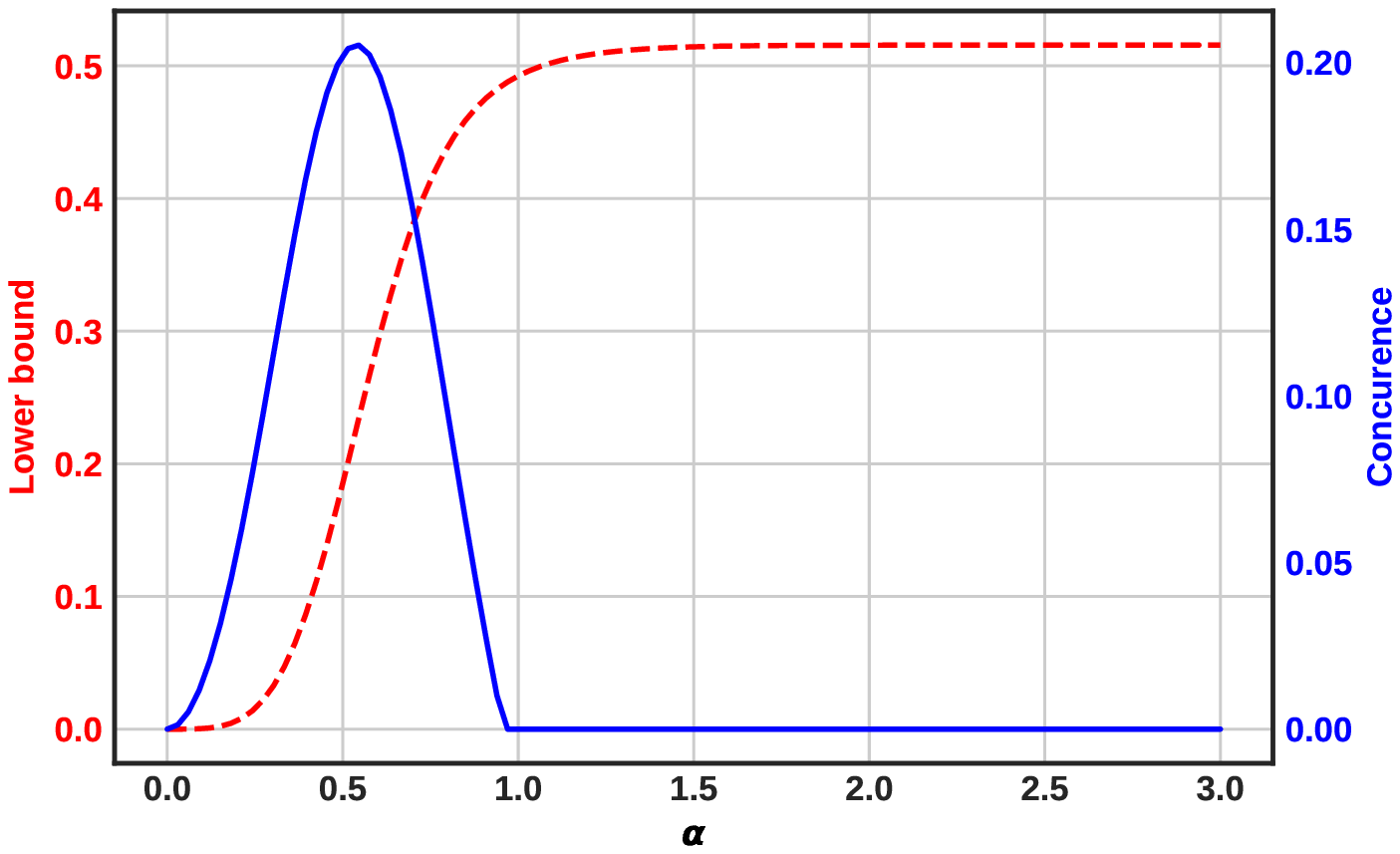}\label{ghzw08}}
\caption{Types of entanglement in mixtures of the  representative states of three qubits with respect to the coherent amplitude $\alpha$ for different values of the mixing parameter $b$. The solid blue line represents the bipartite concurrence and the dashed red line the lower bound for tripartite entanglement.}
    \label{fig:threemixed}
\end{figure}

 Alternatively, with the coherent states, figure \ref{fig:threemixed} shows the types of entanglement present when mixing the representative states based on the coherent state versions  of three qubits classes (\ref{mghzwstate}). Focusing on larger values of $\alpha$ ($\alpha \geq 1.5$) where the states $\arrowvert \alpha \rangle$ and $\arrowvert -\alpha \rangle$ become orthogonal, it is striking that in the case of a balanced mixture $(b=0.5)$, figure \ref{ghzw05} shows that the resulting state belongs to the $[3]_{3}$ class, which means that the $\arrowvert GHZ_{3} \rangle$  is the  "strongest" state. However, in the unbalanced cases \ref{ghzw02} and \ref{ghzw08}  it is seen that, as can be expected, the states with the most weight in the mixture dictates the class in which the mixture falls. As such, the mixture presented in \ref{ghzw02} falls in the $[3]_{2}$ class and the one in \ref{ghzw08} in the $[3]_{3}$ class.

\subsubsection{Four qubits}
 For four qubits, we define a mixture of $\arrowvert GHZ_{4} \rangle$, $\arrowvert W_{4} \rangle$ and $\arrowvert X \rangle$ states as
\begin{equation}
 \rho(b,c) =  a\arrowvert GHZ_{4} \rangle \langle GHZ_{4} \arrowvert + b \arrowvert W_{4} \rangle \langle W_{4} \arrowvert + c \arrowvert X \rangle \langle X \arrowvert,
    \label{mghzwXstate}
\end{equation}
with $a,b,c$ being the mixing parameters ranging from $0$ to $1$ each, such that $a+b+c = 1$.
\begin{figure}[h]
    \centering
    \subfigure[$a=b=c=0.33$]{\includegraphics[width=0.45\linewidth]{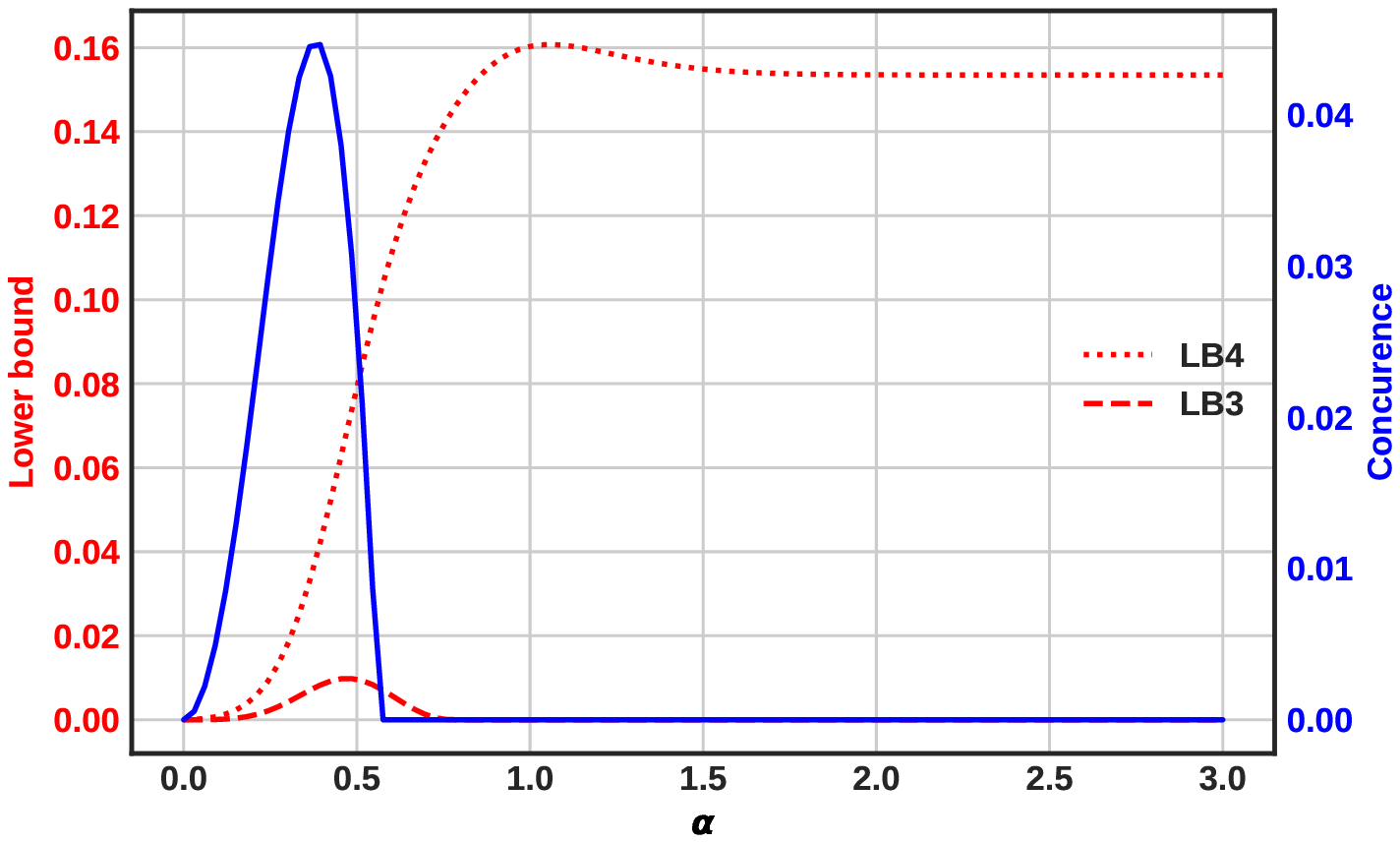}\label{ghzwx3}}
    \subfigure[$a=0.8$ and $b=c=0.1$]{\includegraphics[width=0.45\linewidth]{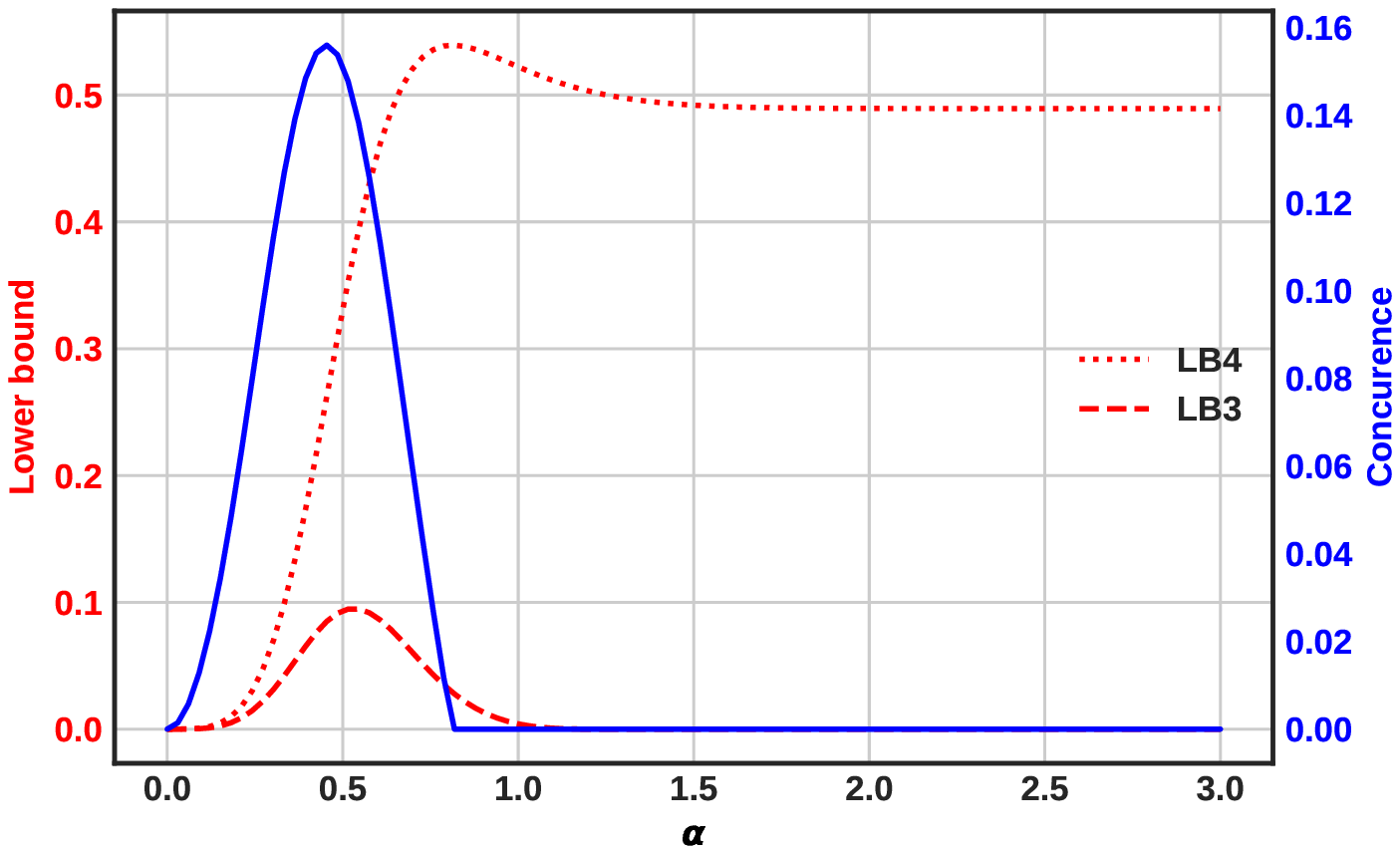}\label{ghzwx811}}
    \subfigure[$b=0.8$ and $a=c=0.1$]{\includegraphics[width=0.45\linewidth]{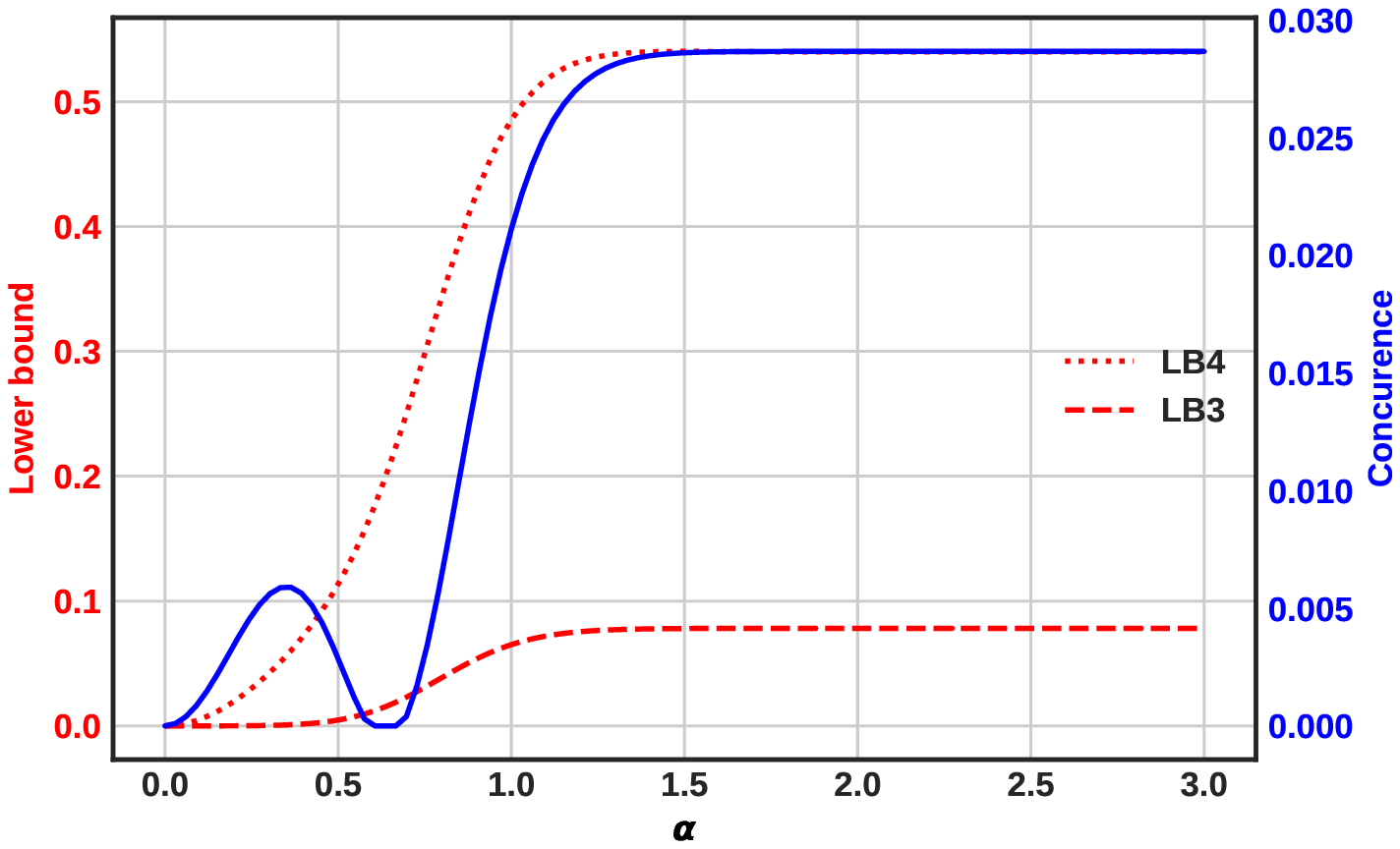}\label{ghzwx181}}
    \subfigure[$c=0.8$ and $a=b=0.1$]{\includegraphics[width=0.45\linewidth]{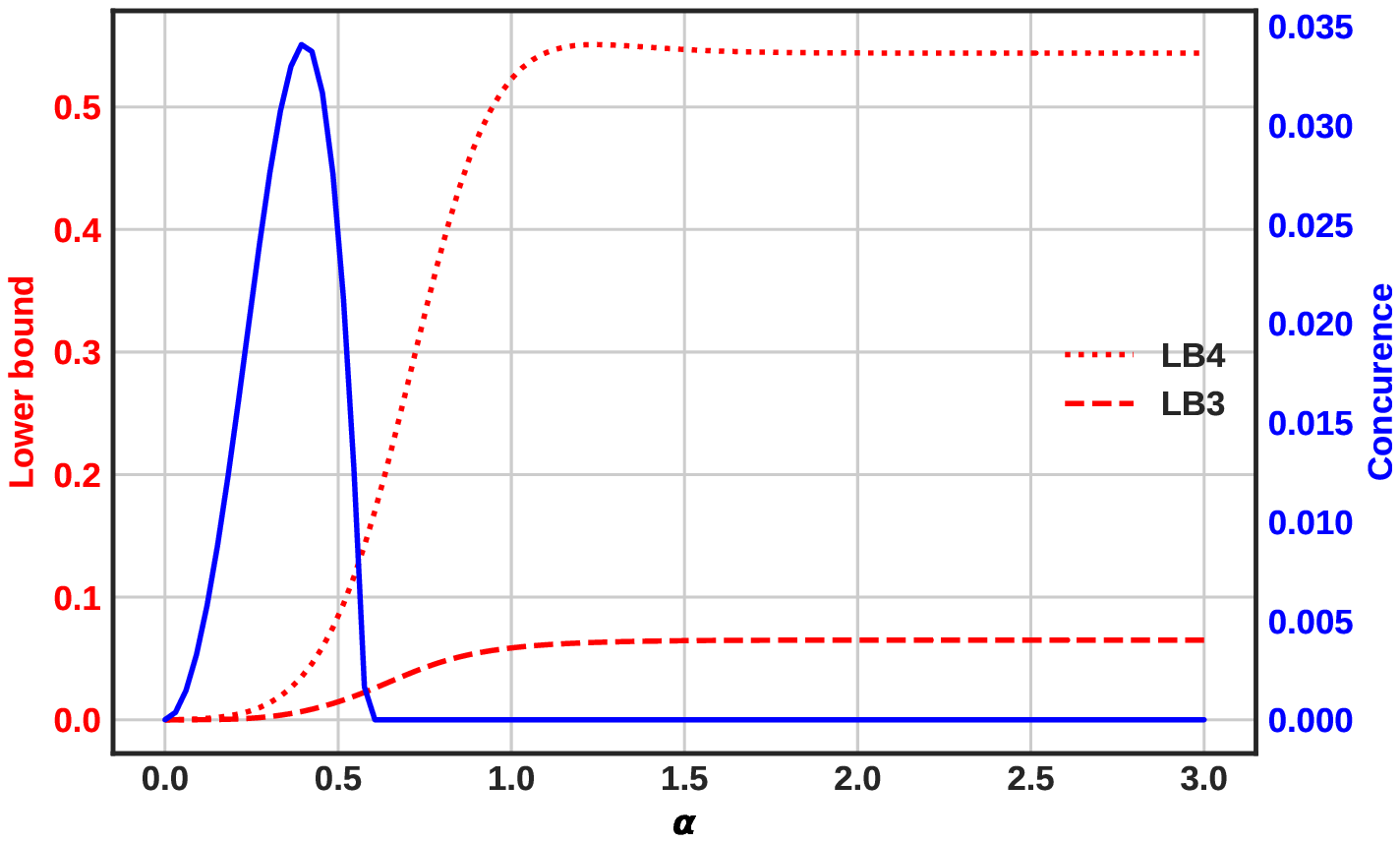}\label{ghzwx118}}
\caption{Types of entanglement in mixed  representative states of Four qubits with respect to the coherent amplitude $\alpha$ for different values of the mixing parameters $a$,$b$ and $c$. The solid blue line represents the bipartite concurrence while the dashed red line the lower bound for tripartite entanglement $LB3$ and the dotted red line the lower bound for four-partite entanglement $LB4$.}
    \label{fig:Fourmixed}
\end{figure}

 Figure \ref{fig:Fourmixed} show the types of entanglement present when mixing the representative states based on the coherent state versions of four qubits classes (\ref{mghzwXstate}). Just like for three qubits, here also, in the case of a balanced mixture $(a=b=c=1/3)$, figure \ref{ghzwx3} shows that the resulting state belongs to the $[4]_{4}$ class, which means that the $\arrowvert GHZ_{4} \rangle$  is the  "strongest" state. However, in the unbalanced cases \ref{ghzwx811}, \ref{ghzwx181} and \ref{ghzwx118} it is seen that, as can be expected, the states with the most weight in the mixture dictates the class in which the mixture falls. As such, the mixture presented in \ref{ghzwx811} falls in the $[4]_{4}$ class and the one in \ref{ghzwx118} in the $[4]_{3}$ class, while the mixture in \ref{ghzwx181} is in the $[4]_{2}$ class. 

Since a similar behavior is noticed in the three qubits case, it can be generalized to an arbitrary number of qubits. Namely that in the unbalanced case, the representative state with most weight dictates the class in which the mixture falls, while the balanced mixture, falls in the $[N]_{N}$ class.

\section{Classification using Machine Learning }
\label{sect.MLcalss}
\subsection{Machine Learning, Deep Learning and Artificial Neural Networks}
Machine learning \cite{Mitchell15ML} is an application of artificial intelligence (AI) that gives systems the ability to automatically learn to do tasks without being explicitly programmed. It focuses on the development of algorithms that can access data to be used to independently learn without any intervention from the programmer. There are three types of machine learning methods: supervised, unsupervised and reinforcement learning.

In the present work, we apply supervised learning methods, in which  the class (output) of each state (input) from a sample, is known, with the goal of finding a function that best maps inputs to outputs and use it to classify new states.

Deep Learning \cite{15} is a subfield of machine learning concerned with algorithms inspired by the structure and function of the human brain called Artificial Neural Networks (ANN). Each "Neuron", called perceptron in Artificial intelligent (AI), is a mathematical function that collects and classifies information according to a particular structure.

 In multi-layer perceptron (MLP), the perceptrons are arranged in interconnected layers. An input layer (sensory unit) collects the input patterns while an output layer (response unit) contains classifications or exit value to which input patterns may be assigned. In between, hidden layers (associator unit) adjust input weights until the neural network loss is minimal. It is assumed that the hidden layers capture the salient features of the input data that have predictive power with respect to the corresponding output. In our case, the features to be extracted are the density matrix elements or their combinations that might be responsible for each entanglement class.

\subsection{Data set}
The procedure to follow in order to prepare the data set, depends on the dimension of the system. In the following we outline the procedure depending on the overall number of qubits considered.

\noindent \textbf{Bipartite states:}

\begin{enumerate}
	\item Draw a random density matrix state $ \rho $ of 2 qubits.
	\item Compute the concurrence of $ \rho $, if it is equal to 0 we save $\rho  $ in our data set as "Separable", else we save it as "Entangled".
\end{enumerate}

\noindent \textbf{Multipartite states:}

\begin{enumerate}
	
	 \item Draw a random  density matrix state $ \rho $ according to the number of qubits (3 or 4 in our applications).

	\item Compute the different $i$-tangles  and assigne the corresponding class according to the classfication established in \ref{sbsect_tree} and \ref{sbsect_class}. Namely for 3 and 4 qubits cases we proceed as follows.
	\begin{itemize}
	    \item For 3 qubits, if $ \tau_{1} = 0 $, we save $\rho  $ in our data set as "separable", else if $ \tau_{2} \neq 0 $, we save it in our data set as "$ [3]_{2} $", else we save it as "$[3]_{3} $".
	    
	    \item For 4 qubits, if $ \tau_{1} = 0 $ , we save $\rho  $ in our data set as "separable", else if $ \tau_{2} \neq 0 $ , we save it in our data set as "$[4]_{2} $", else if $ \tau_{3} \neq 0 $ we save it as "$[4]_{3}$", else we save it as "$[4]_{4}$".
	\end{itemize}
\end{enumerate}

The final dataset we generated consists of 540 000 density matrices scattered as follows
\begin{enumerate}
\item[$\square$] for 2-qubit systems, we have 400 000 density matrices, 200 000 pure and 200 000 mixed , each type contain 100 000 of each class \textit{separable/entangled}.

\item[$\square$]  for 3-qubit systems, we have 60 000 density matrices, 30 000 pure and 30 000 mixed, each type contain 10 000 of each class \textit{k-separable or separable} /$ [3]_{2} $/$[3]_{3} $.

\item[$\square$] for 4-qubit systems, we have 80 000 density matrices, 40 000 pure and 40 000 mixed, each type contain 10 000 of each class \textit{k-separable or separable} /$ [4]_{2} $/ $[4]_{3} $/ $[4]_{4} $.
\end{enumerate}

The discrepancy in the number of data generated is mainly due to the increase in the number of elements to be generated accompanying the increase of number of q-bits. For instance, for a system of 2-qubits only $2^2\times 2^2=16$ elements are needed while for 4-qubits, one needs $2^4\times 2^4=256$ elements, all while imposing the density matrices conditions. It turns out that some classes are easier (quicker) to generate than others, however to equilibrate things we chose equal numbers of each class generated.

This data was genarated through computational resources of HPC-MARWAN (hpc.marwan.ma) provided by the National Center for Scientific and Technical Research (CNRST) , Rabat, Morocco

\subsection{Results}
After preparing the data, the first task is training the ANN classifiers to distinguish the different classes. To assess the overall performance of our classifier, we remove the redundancy to ensure that the training and testing data are totally different, and we split the data set in two parts:
80$ \%  $ for training and  20$ \%  $ for testing. The goal of this splitting is to test our model with new
information (testing data), so we can ensure the model’s robustness against over-fitting, among other things. One can state that a given model is actually good, when it achieves a high performance in the testing phase.

To display the performance of our classifiers we use the error matrices, also known as confusion matrices \cite{20}, as they make it easy to see how much  the system is confusing two classes. The confusion matrices are two dimensional tables \textit{i.e} $(N,N)$ matrices  with $N$ being the number of classes, such that one dimension represents the true class, and the other dimension represents the predicted class. The diagonal elements of these matrices represent the number of samples belonging to the correct class (successful prediction) while the other elements represent the number of samples for which the predicted class is wrong.

There are different measures, based on the confusion matrix, that allow to assess the performance of a classification. In the following, we will use the $Accuracy$ and the $Precision_i$ of a given class $i$, defined hereafter.
\begin{enumerate}
\item Accuracy is the proportion of the total number of correct predictions:
\begin{equation}
    Accuracy=\frac{\sum_{i-1}^{N} M_{ii}}{ \sum_{i-1}^{N} \sum_{j=1}^{N} M_{ij}}.
    \label{Acc}
\end{equation}

\item Precision is a measure of the number of cases in which  a specific class $i$ has been correctly predicted:
\begin{equation}
    Precision_{i}=\frac{M_{ii}}{ \sum_{k=1}^{N} M_{ki}}.
    \label{prec}
\end{equation}
\end{enumerate}
In the above definitions, $M_{ij}$ are the confusion matrix elements and $N$ is the number of classes.

\begin{figure}[h]
    \centering
    \subfigure[Pure]{\includegraphics[width=0.49\linewidth]{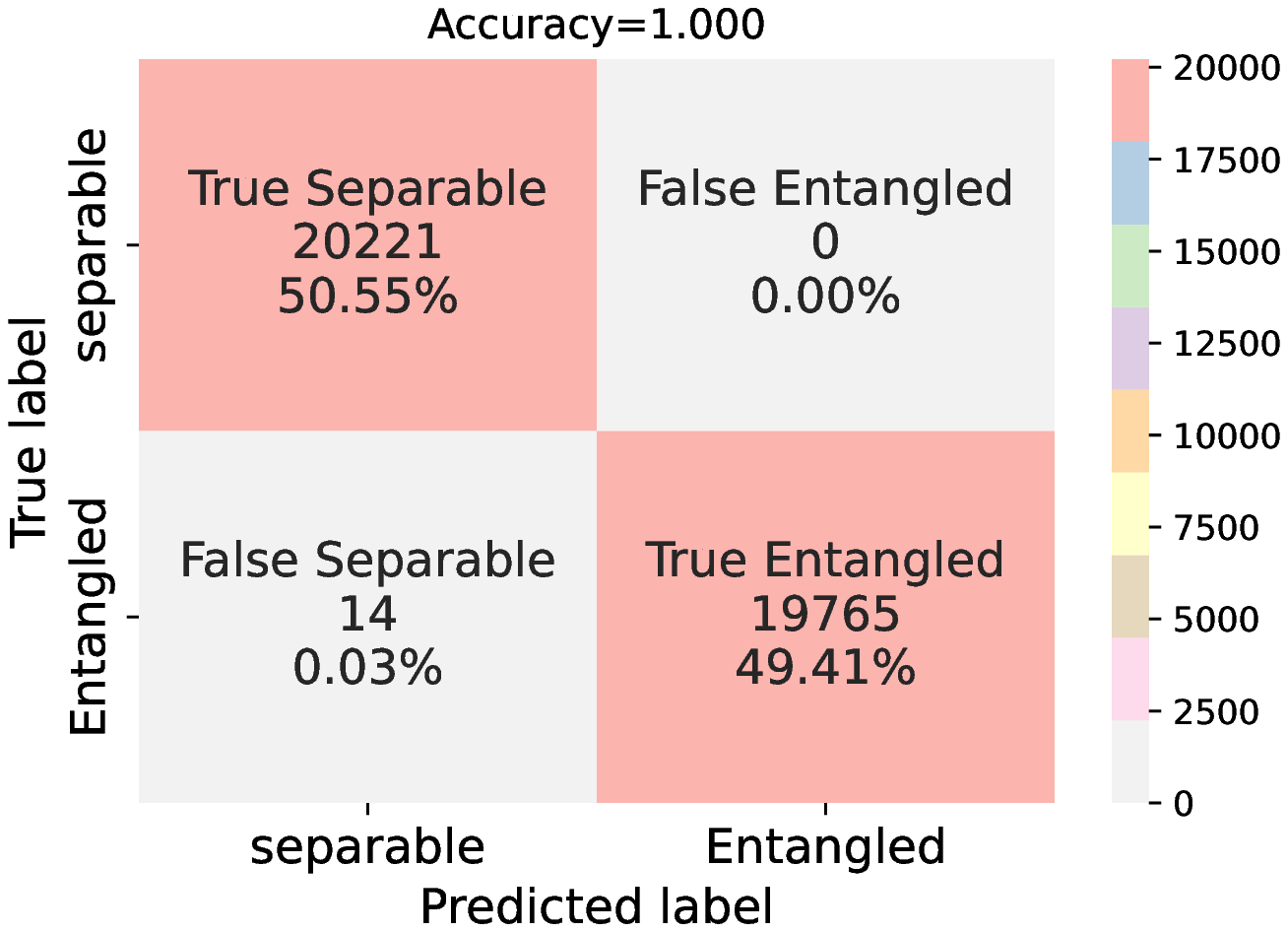}\label{conf2p}}
    \subfigure[Mixed]{\includegraphics[width=0.49\linewidth]{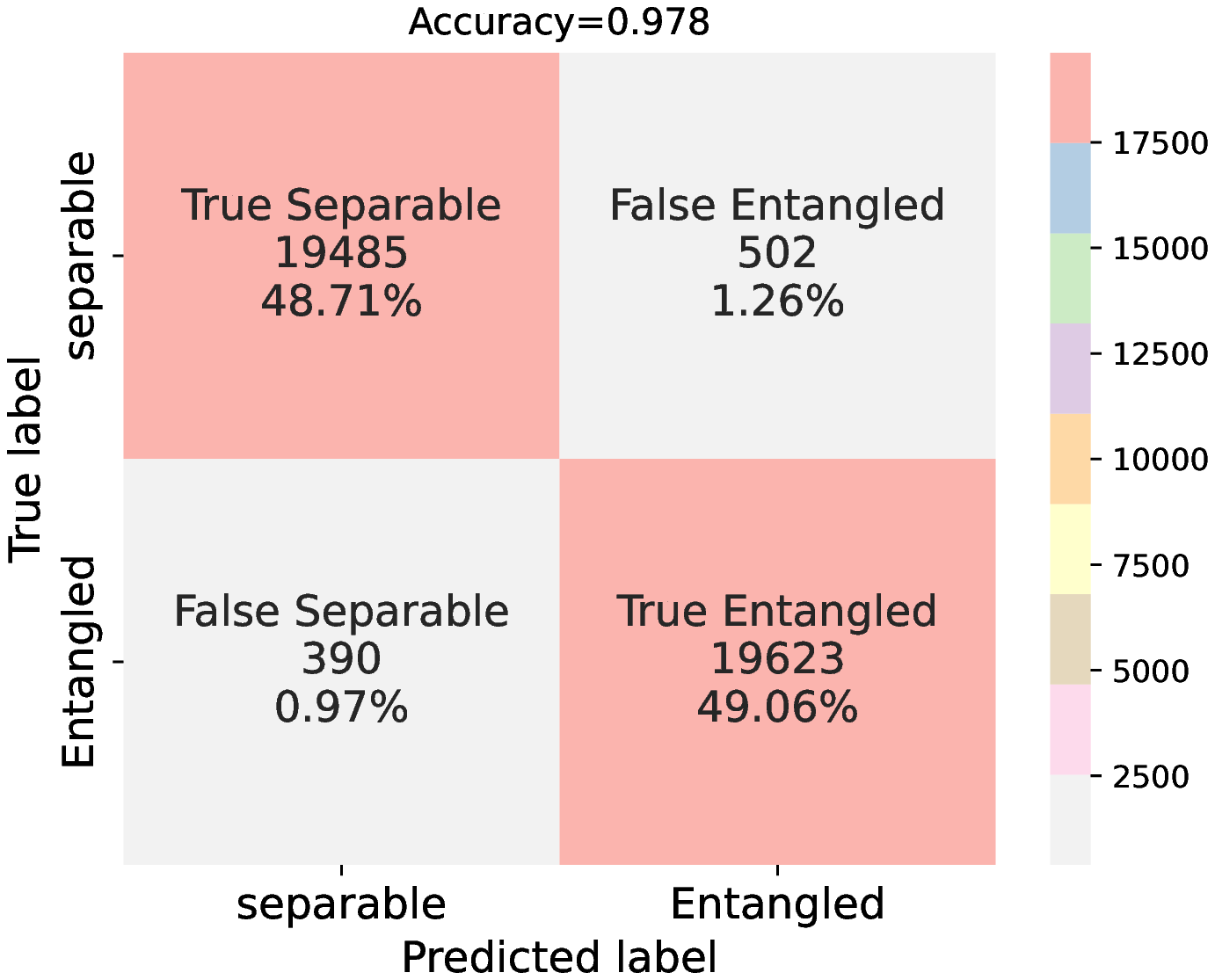}\label{conf2m}}
\caption{The testing confusion matrix for the classification of pure (a) and mixed  (b) bipartite   density matrices.}
    \label{fig:2cf}
\end{figure}

Figure \ref{fig:2cf} represents the testing confusion matrices of our classifier for pure \ref{conf2p} and mixed \ref{conf2m} bipartite states, where the rows correspond to the true classes (Entangled or Separable) and the columns to the predicted classes. As stated before, the correctly classified states are accounted for in the diagonal entries and the incorrectly classified ones in the off-diagonal, with the number of states and their proportion, in each case, being reported. This confusion matrices  shows that our classifier achieved an overall performance, as estimated using equation (\ref{Acc}), that is almost perfect for pure states and 97.8$\%$ for mixed states. In the case of pure states, the entangled states are accurately classified with a precision of 99.92$\%$  and the separable ones with a precision of 100$\%$, but for mixed states, the entangled ones are accurately classified with a precision of 98.05$\%$ and the separable states with a precision of 97.48$\%$ as calculated using equation (\ref{prec}).

\begin{figure}[h]
    \centering
    \subfigure[Pure]{\includegraphics[width=0.49\linewidth]{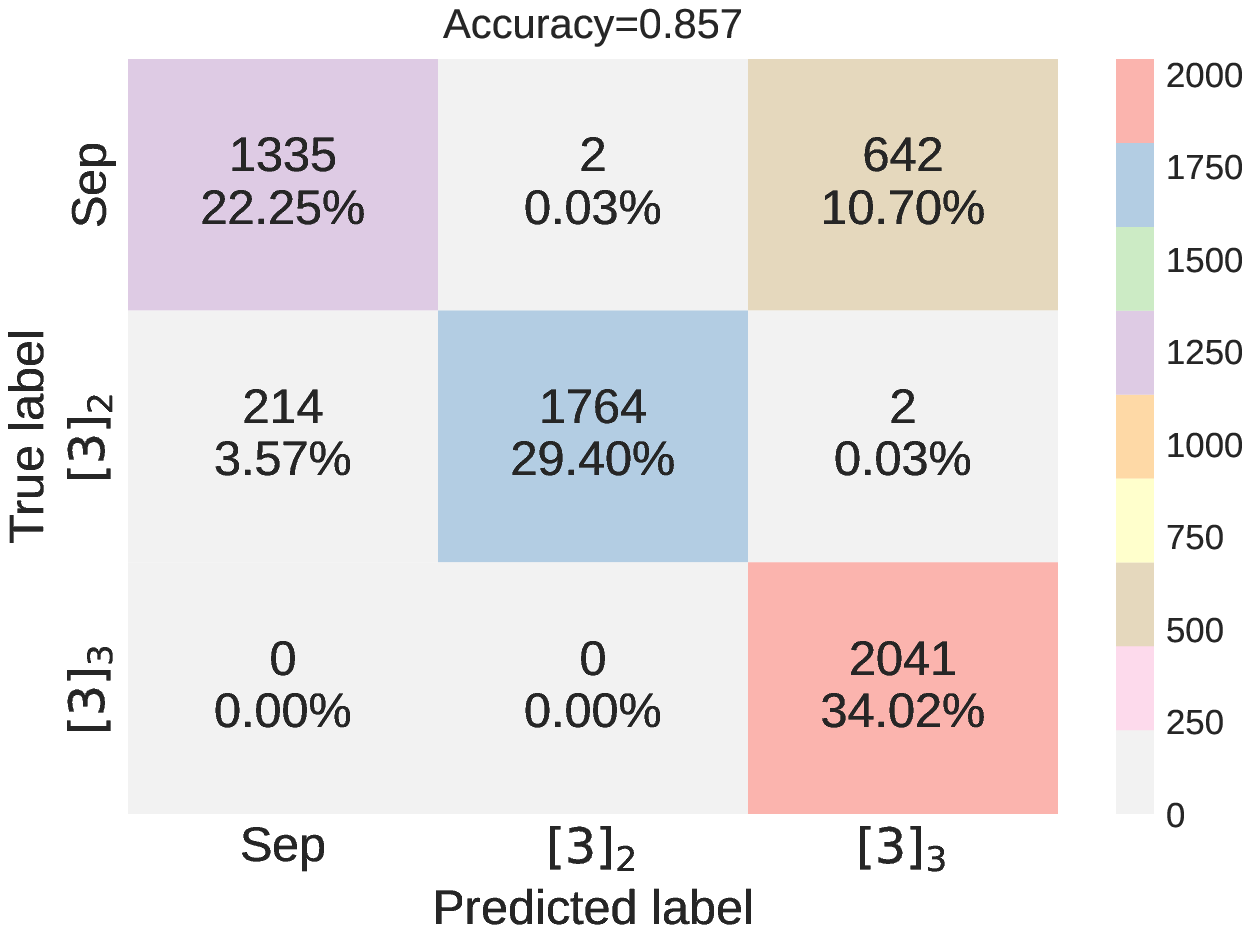}\label{conf3p}}
    \subfigure[Mixed]{\includegraphics[width=0.49\linewidth]{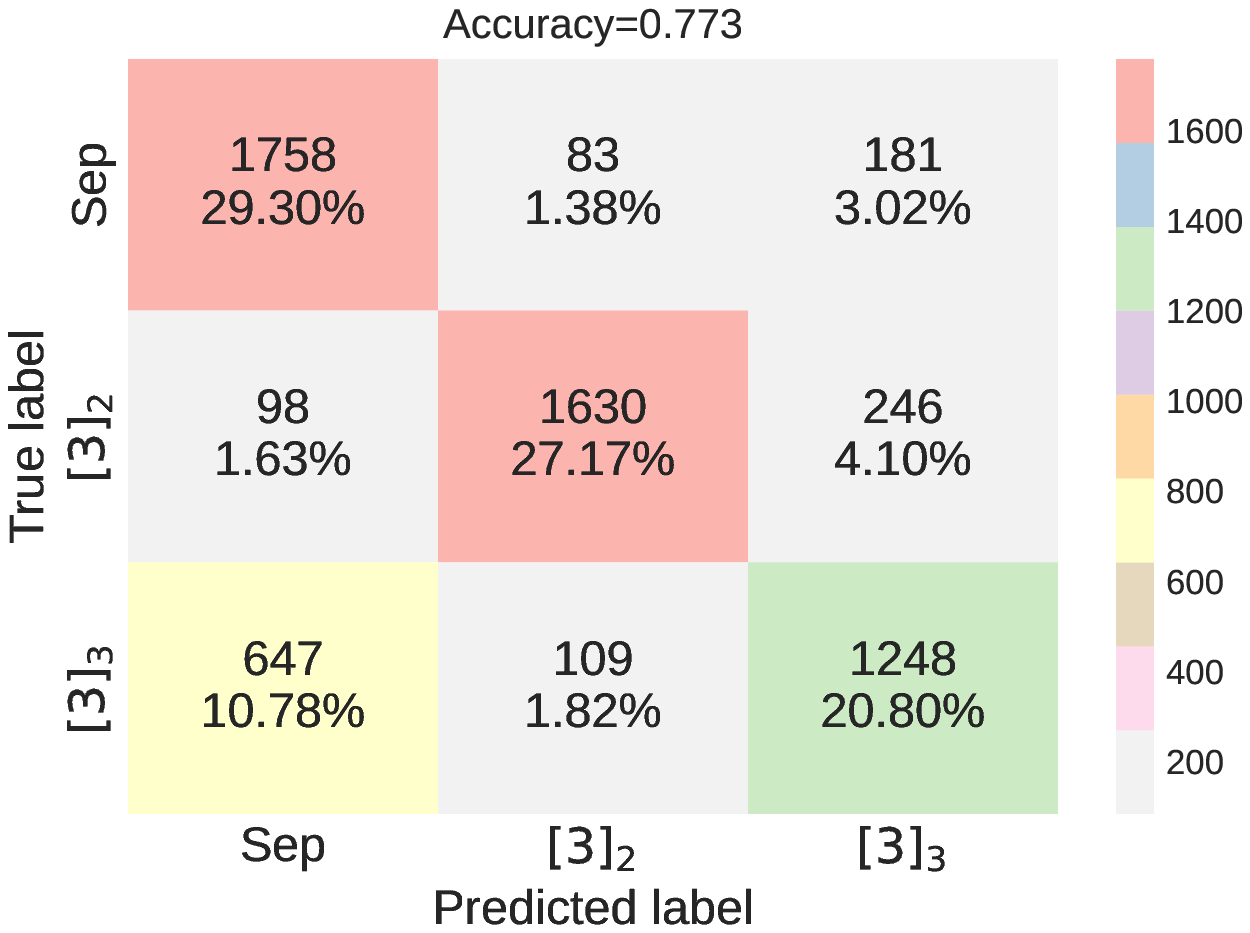}\label{conf3m}}
\caption{The testing confusion matrix for the classification of pure (a) and mixed (b) tripartite   density matrices.}
    \label{fig:3cf}
\end{figure}

Figure \ref{fig:3cf} represents the testing confusion matrices of our classifier for pure \ref{conf3p} and mixed \ref{conf3m} tripartite states, where again the columns correspond to the predicted classes (k-separable/separable, $[3]_{2} $ or $[3]_{3}$) and the rows to the corresponding true classes. This confusion matrices  shows that our classification achieved an overall performance of 85.7$\%$ for pure systems and 77.3$\%$ for mixed systems, where it accurately classified separable/$k$-separable states with precision of 86.18$\%$ for pure systems and 86.94$\%$ for mixed systems, $[3]_{2}$ states with 89.09$\%$ for pure systems and 82.57$\%$ for mixed systems, and $[3]_{3}$ states with a precision of 100$\%$ for pure systems and 62.27$\%$  for mixed systems.

\begin{figure}[h]
    \centering
    \subfigure[Pure]{\includegraphics[width=0.49\linewidth]{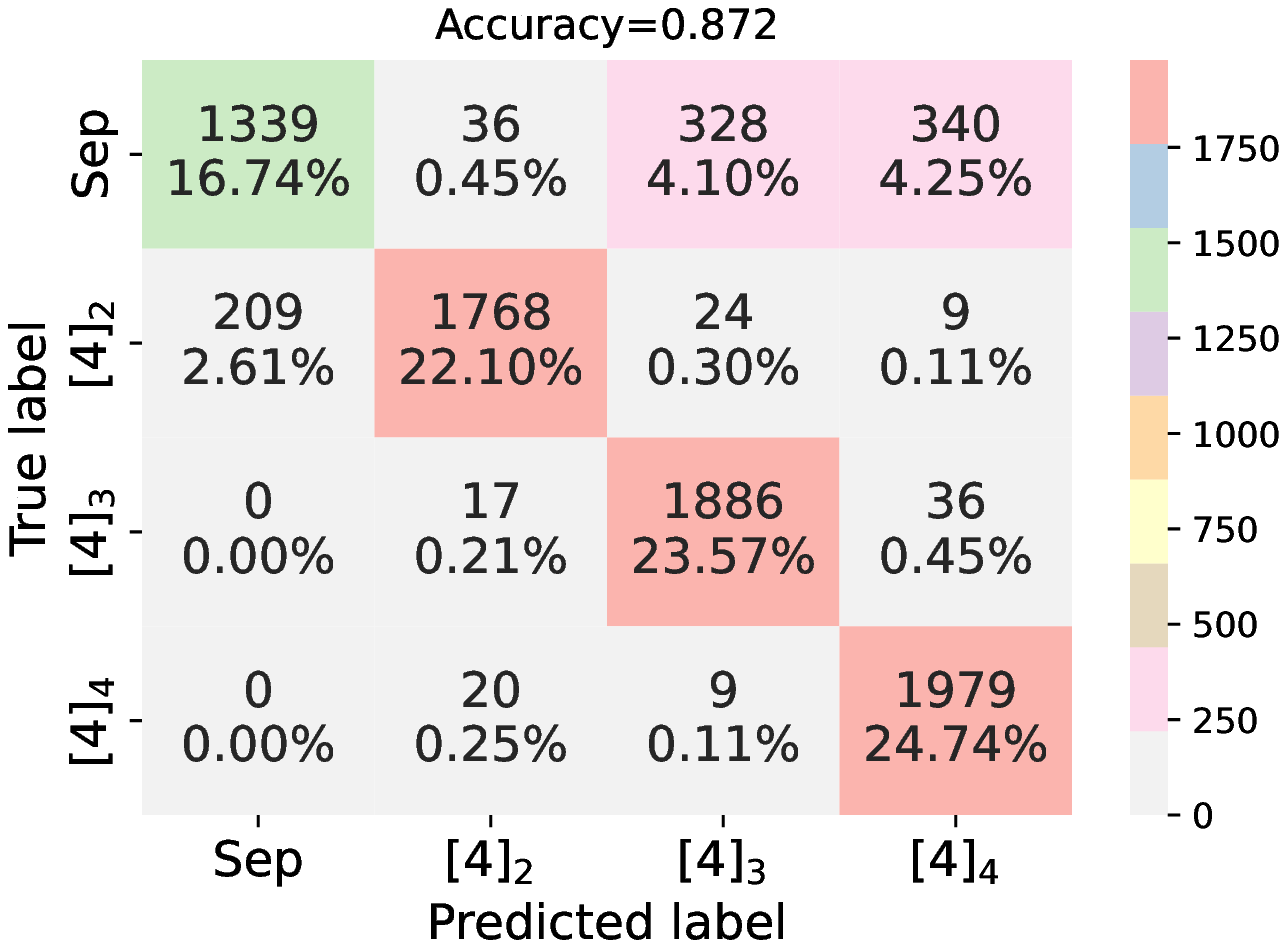}\label{conf4p}}
    \subfigure[Mixed]{\includegraphics[width=0.49\linewidth]{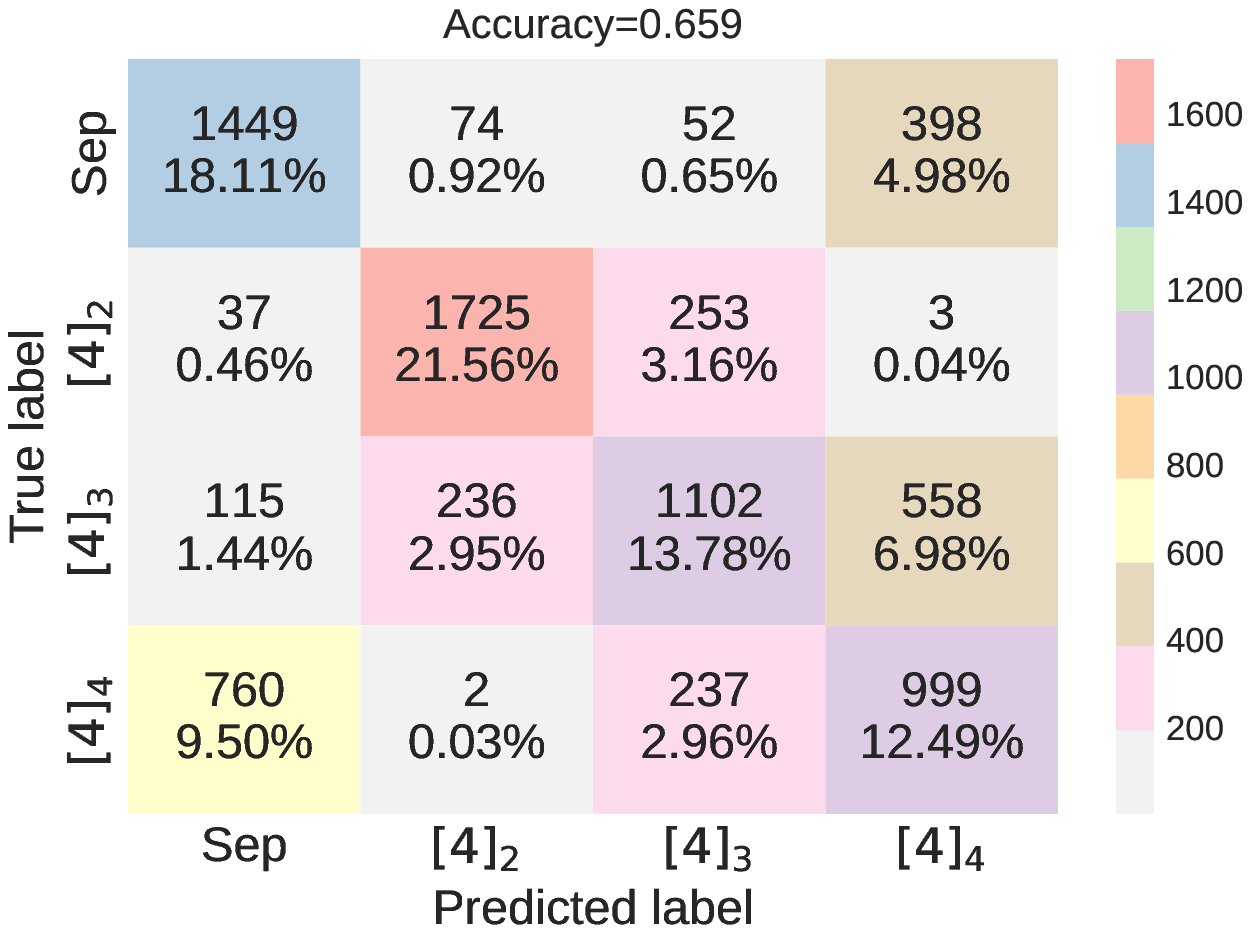}\label{conf4m}}
\caption{The testing confusion matrix for the classification of pure (a) and mixed (b) fourpartite   density matrices.}
    \label{fig:4cf}
\end{figure}

Figure \ref{fig:4cf} represents the testing Confusion Matrices of our classifier for pure \ref{conf4p} and mixed \ref{conf4m} fourpartite states where the classes being distinguished are the separable/$k$-separable vs. $[4]_{4} $  vs. $[4]_{3} $  vs. $[4]_{2} $. As in the previous cases, the  number of states as well as their proportion, in each case, are reported in the entries of the matrix. This confusion matrices shows that our classifier achieved an overall performance of 87.2$\%$ for pure state and 65.9$\%$ for mixed ones. The separable/$k$-separable states  are accurately classified with a precision of 86.49$\%$ for pure states and 73.44$\%$ for mixed states, the $[4]_{2} $ states with   96.03$\%$ for pure sysytems and 85.48$\%$ for mixed systems, the $[4]_{3} $ states with a precision of 83.93$\%$ for pure systems and 54.79$\%$ for mixed states, and the $[4]_{4} $ states with a precision of 83.71$\%$ for pure states and 50$\%$ for mixed states.

To test the importance of using Machine learning in quantum information area, we must test our classifier with some known separable/$k$-separable states as well as the representative states of the different classes such as the four Bell states for $2$-qubits, $\arrowvert GHZ_3 \rangle$ and $\arrowvert W_3 \rangle$ states for $3$-qubits (shown in the figures \ref{threeW}, \ref{threeGHZ}), and $\arrowvert GHZ_4 \rangle$ , $\arrowvert W_4 \rangle$ , $\arrowvert X_4 \rangle$  states in the $4$-qubits case (see figures \ref{FourGHZ}, \ref{FourW} and \ref{FourX} respectively). Mixtures of the said representative states must also be included in the test, to discuss its performance with mixed states. The results of this test are summarized in Table \ref{tableau1}.

It is seen in Tables \ref{tableau1} and \ref{tableau2} that our model distinguishes perfectly the different classes of the different well-known states. The errors appearing in the tables are all stemming from elements of the confusion matrix with a high error percentage. The results in the tables are in agreement with the the results of the confusion matrices (Figures \ref{fig:2cf}, \ref{fig:3cf} and \ref{fig:4cf}). One clue that transpires through Tables \ref{tableau1} and \ref{tableau2} is that most errors occur in cases where the number and location of non-zero matrix elements of the tested state is close to that of the well established states (like the representative states) while still belonging to different classes.

\begin {table}[h]
\begin{center}

\begin{tabular}{ |l|l|l|l| }
	\hline
	\multicolumn{4}{ |c| }{verification of pure states } \\
	\hline
	Dim & state & True class & Predict class \\ \hline
	\multirow{3}{*}{2 qubits} & $\arrowvert \psi_{1} \rangle = \dfrac{1}{\sqrt{2}}( \arrowvert 00 \rangle + \arrowvert 11 \rangle)$ & Ent & Ent $\checkmark$ \\
	& $\arrowvert \psi_{2} \rangle = \dfrac{1}{\sqrt{2}}( \arrowvert 01 \rangle + \arrowvert 10 \rangle)$ & Ent & Ent $\checkmark$ \\
	& $\arrowvert \psi_{3} \rangle = \dfrac{1}{\sqrt{2}}( \arrowvert 00 \rangle - \arrowvert 11 \rangle)$ & Ent & Ent $\checkmark$\\
	& $\arrowvert \psi_{4} \rangle = \dfrac{1}{\sqrt{2}}( \arrowvert 01 \rangle - \arrowvert 10 \rangle)$ & Ent & Ent $\checkmark$\\
	& $\arrowvert \psi_{5} \rangle =\dfrac{1}{\sqrt{2}}( \arrowvert 00 \rangle + \arrowvert 01 \rangle)$ & Sep & Sep $\checkmark$\\
	& $\arrowvert \psi_{6} \rangle =\dfrac{1}{\sqrt{2}}( \arrowvert 10 \rangle + \arrowvert 11 \rangle)$ & Sep & sep $\checkmark$ \\ 
	& $\arrowvert \psi_{7} \rangle =\dfrac{1}{2}( \arrowvert 00 \rangle + \arrowvert 01 \rangle + \arrowvert 10 \rangle + \arrowvert 11 \rangle)$ & Sep & Sep $\checkmark$ \\ 
	\hline
	\multirow{4}{*}{3 qubits} & $\arrowvert \phi_{1} \rangle =\dfrac{1}{\sqrt{2}}( \arrowvert 000 \rangle + \arrowvert 111 \rangle)$ & $[3]_{3}$ &  $[3]_{3}$  $\checkmark$ \\
	& $\arrowvert \phi_{2} \rangle =\dfrac{1}{\sqrt{3}}( \arrowvert 001 \rangle + \arrowvert 010 \rangle+ \arrowvert 100 \rangle)$ & $[3]_{2}$ & $[3]_{2}$ $\checkmark$\\
	& $\arrowvert \phi_{3} \rangle =\dfrac{1}{\sqrt{3}}( \arrowvert 110 \rangle + \arrowvert 101 \rangle+ \arrowvert 011 \rangle)$ & $[3]_{2}$ & $[3]_{2}$ $\checkmark$ \\
	& $\arrowvert \phi_{4} \rangle =\dfrac{1}{\sqrt{3}}( \arrowvert 010 \rangle + \arrowvert 001 \rangle+ \arrowvert 011 \rangle)$ & K-Sep & Sep $\checkmark$ \\ 
	& $\arrowvert \phi_{5} \rangle = \dfrac{1}{\sqrt{2}}( \arrowvert 000 \rangle + \arrowvert 001 \rangle)$ & Sep & $[3]_{3}$ \xmark  \\
	 \hline
	\multirow{5}{*}{4 qubits} & $\arrowvert \chi_{1} \rangle = \dfrac{1}{\sqrt{2}}( \arrowvert 0000 \rangle + \arrowvert 1111 \rangle)$ & $[4]_{4}$ & $[4]_{4}$ $\checkmark$ \\
	& $\arrowvert \chi_{2} \rangle = \dfrac{1}{2}( \arrowvert 0001 \rangle + \arrowvert 0010 \rangle+ \arrowvert 0100 \rangle+\arrowvert 1000 \rangle)$ & $[4]_{2}$& $[4]_{2}$ $\checkmark$ \\
	& $ \arrowvert \chi_{3} \rangle = \dfrac{1}{2}( \arrowvert 1110 \rangle + \arrowvert 1101 \rangle+ \arrowvert 1011 \rangle + \arrowvert 0111 \rangle)$ & $[4]_{2}$ & Sep \xmark \\
	& $\arrowvert \chi_{4} \rangle = \dfrac{1}{\sqrt{5}}( \arrowvert 0000 \rangle + \arrowvert 0111 \rangle+ \arrowvert 1011 \rangle + \arrowvert 1101 \rangle + \arrowvert 1110 \rangle)$ & $[4]_{3}$ & $[4]_{3}$ $\checkmark$ \\ 
	& $\arrowvert \chi_{5} \rangle =\dfrac{1}{\sqrt{5}}( \arrowvert 1000 \rangle + \arrowvert 0100 \rangle+ \arrowvert 0010 \rangle + \arrowvert 0001 \rangle + \arrowvert 1111 \rangle)$ & $[4]_{3}$ & $[4]_{3}$ $\checkmark$ \\ 
	& $\arrowvert \chi_{6} \rangle = \dfrac{1}{2}( \arrowvert 0000 \rangle + \arrowvert 0011 \rangle+ \arrowvert 0010 \rangle + \arrowvert 0001 \rangle)$ & K-Sep & Sep $\checkmark$ 
	\\
	 \hline

\end{tabular}
\caption{\label{tableau1}Verification of the classification of some well known pure states.}
\end{center}
\end{table}

\newpage

\begin {table}[H]
\begin{center}

\begin{tabular}{ |l|l|l|l| }
	\hline
	\multicolumn{4}{ |c| }{verification of mixed states} \\
	\hline
	Dim & state & True class & Predict class \\ \hline
	\multirow{3}{*}{2 qubits} & 
	$\rho_{1}  = 0.5(\arrowvert \psi_{1} \rangle \langle \psi_{1}  \arrowvert) +0.5( \arrowvert \psi_{3} \rangle \langle \psi_{3}  \arrowvert )$ & Sep & Sep $\checkmark$ 
    \\
	& $ \rho_{2}  = 0.8(\arrowvert \psi_{1} \rangle \langle \psi_{1}  \arrowvert) +0.2( \arrowvert \psi_{3} \rangle \langle \psi_{3}  \arrowvert )$ & Ent & Ent
	$\checkmark$ 
	\\
	& $ \rho_{3}  = 0.5(\arrowvert \psi_{2} \rangle \langle \psi_{2}  \arrowvert) +0.5( \arrowvert \psi_{6} \rangle \langle \psi_{6}  \arrowvert )$ & Sep & Sep
	$\checkmark$ 
	\\
	& $ \rho_{4}  = 0.8(\arrowvert \psi_{2} \rangle \langle \psi_{2}  \arrowvert) +0.2( \arrowvert \psi_{6} \rangle \langle \psi_{6}  \arrowvert )$ & Ent & Ent
	$\checkmark$ 
	\\
	\hline
	\multirow{4}{*}{3 qubits} & 
	 $ \sigma_{1}  = 0.8(\arrowvert \phi_{1} \rangle \langle \phi_{1}  \arrowvert) +0.2( \arrowvert \phi_{3} \rangle \langle \phi_{3}  \arrowvert )$ & $[3]_{3}$ & $[3]_{3}$
	$\checkmark$ 
	\\
	& $ \sigma_{2}  = 0.8(\arrowvert \phi_{1} \rangle \langle \phi_{1}  \arrowvert) +0.2( \arrowvert \phi_{3} \rangle \langle \phi_{3}  \arrowvert )$ & $[3]_{3}$ & $[3]_{3}$
	$\checkmark$ 
	\\
	& $ \sigma_{3}  = 0.8(\arrowvert \phi_{1} \rangle \langle \phi_{1}  \arrowvert) +0.2( \arrowvert \phi_{3} \rangle \langle \phi_{3}  \arrowvert )$ & $[3]_{3}$ & $[3]_{3}$
	$\checkmark$ 
	\\
	& $ \sigma_{4}  = 0.8(\arrowvert \phi_{1} \rangle \langle \phi_{1}  \arrowvert) +0.2( \arrowvert \phi_{3} \rangle \langle \phi_{3}  \arrowvert )$ & $[3]_{3}$ & Sep
	\xmark 
	\\
	 \hline
	\multirow{5}{*}{4 qubits} 
	& $ \mu_{1}  = 0.5(\arrowvert \chi_{1} \rangle \langle \chi_{1}  \arrowvert) +0.5( \arrowvert \chi_{3} \rangle \langle \chi_{3}  \arrowvert )$ & $[4]_{4}$ & $[4]_{4}$ $\checkmark$
	\\
	& $ \mu_{2}  = 0.2(\arrowvert \chi_{2} \rangle \langle \chi_{2}  \arrowvert) +0.8( \arrowvert \chi_{4} \rangle \langle \chi_{4}  \arrowvert )$ & $[4]_{3}$ & $[4]_{3}$ $\checkmark$
		\\
	& $ \mu_{3}  = 0.8(\arrowvert \chi_{6} \rangle \langle \chi_{6}  \arrowvert) +0.2( \arrowvert \chi_{5} \rangle \langle \chi_{5}  \arrowvert )$ & Sep & Sep $\checkmark$
\\
	& $ \mu_{4}  = 0.8(\arrowvert \chi_{1} \rangle \langle \chi_{1}  \arrowvert) +0.2( \arrowvert \chi_{3} \rangle \langle \chi_{3}  \arrowvert )$ & $[4]_{4}$ & Sep
	\xmark\\
& $ \mu_{5}  = 0.8(\arrowvert \chi_{5} \rangle \langle \chi_{5}  \arrowvert) +0.2( \arrowvert \chi_{2} \rangle \langle \chi_{2}  \arrowvert )$ & $[4]_{3}$ & $[4]_{4}$
	\xmark\\
	 \hline
\end{tabular}
\caption{\label{tableau2}Verification of the classification of some mixture of  the states used in Table 1.}
\end{center}
\end{table}

\section{Discussion and Conclusion}
In summary, we have introduced a new classification for multipartite fully entangled states. The classification is based on the robustness of entanglement against partial trace operations.

We have used a neural networks model to automatically classify random quantum states , without the need to explicitly calculate the corresponding entanglement measures each time. Instead, the NN model learns to distinguish the different classes directly from the density matrix elements.

We applied the technique to classify states of two, three and four qubits, into separable (or $k$-separable) and fully entangled states while specifying the corresponding class denoted $[N]_j$, where the index $j=1,2,\dots , N-1$, specifies the actual class given the number of qubits $N$.

The proposed algorithm can be integrated in any quantum information processing protocol that rely on a specific type of entanglement or, for that matter, adapt depending on the class of multipartite entanglement detected. As a matter of fact, The importance of any classification scheme stems from the usefulness of the resulting types of multipartite entanglement in the targeted applications. As an illustration for this, let us consider a given teleportation protocol involving three parties (Alice and Bob being the primary parties and Charlie being an ancilla) that share an entangled state. When Alice teleports an arbitrary qubit to Bob, if she decides to break the operation in the case Charlie measures his qubit, then they need to use a  $ [ 3 ]_{3} $ class state.  Otherwise, if Alice wants to  continue this teleportation independently of Charlie's qubit, they'll need to use a  $ [ 2 ]_{3} $ class state.

Following alternative lines of thought, a quantum teleportation protocol was introduced in \cite{17} and a comparison of the efficiency of different 4-qubits states to  teleport two qubits resulted in the $GHZ$ state (belonging to the $[ 4 ]_{4} $ class) being more efficient than the $W$ state which belongs to the $ [ 2 ]_{4} $ class. Notwithstanding the foregoing, \cite{18} establishes that a 4-qubit $GHZ$ state class is not a good candidate for a perfect quantum teleportation using their protocol and the so-called $\chi$ state \cite{qsistate}, which fall in the class $[ 3 ]_4 $ of our classification, provides a better channel.

On the other hand, for quantum radars \cite{QRadar} making use of fully-entangled particles, whereby one part is kept in the system while the other parties are sent towards a target. So when one of the sent particles is traced out (interacting with the environment), the loss of entanglement with the other particles depends on the class of entangled state chosen in the first place. Bearing this in mind, it is evident that while states with genuine multipartite entanglement ($N$-entanglement) will have the higher efficiency, they will be more vulnerable to the interaction with environment. So in this case an optimisation procedure is necessary depending on the types of interactions in play.

Other results, suggest that multipartite entanglement speeds up quantum key distribution in networks \cite{QKDMult} reaching different performances with different classes of entanglement.

The results of this paper can be exploited in other directions as well. Among these, work is in progress to implement this technique to detect the entanglement transitions from one class to another in dynamical systems. It will be interesting to investigate the generalisation to larger dimensions and deal with qudits instead of qubits. This is even so important as the existence of a proper measure of entanglement even in the bipartite case is still debated. 

As a final note, it is believed that by using quantum machine learning algorithms will be more efficient, especially in the cases where the accuracy of the current technique was relatively small (like in the case of fourpartite mixed states here). This belief comes from the fact, that quantum algorithms should recognize quantum features (like entanglement) more "intuitively", thus resulting in more accurate classification schemes \cite{superQMLshuld}. However, this last technique is still limited by the number of qubits that one can use in the actual quantum computers.

\newpage

\begin{appendices}
\section{Entanglement measures}
\label{appendix1}

\begin{itemize}
\item Wooter's concurrence \cite{11}:
\begin{equation}
	C_{i/j} = max \{ 0,\lambda(1)-\lambda(2)-\lambda(3)-\lambda(4)\}
	\label{Wconc}
\end{equation}

 In which $\lambda(i)$ are the eigenvalues, in decreasing order, of the Hermitian matrix $\sqrt{\sqrt{\rho}\tilde{\rho}\sqrt{\rho}}$, with $\tilde{\rho} = (\sigma_{y} \otimes \sigma_{y}) \rho^{*} (\sigma_{y} \otimes \sigma_{y})$. \\
\item The I-concurrence \cite{10}:
\begin{equation}
	C^{p} = C_{i/j} = \sqrt{2-Tr(\rho_{i})^{2}-Tr(\rho_{j})^{2}}
\label{iconc}	
\end{equation}

with $\rho_{i}$ and $\rho_{j}$  are respectively the reduced density matrices of bipartition $i$ and $j$

\item The genuin tangle of pure three partite systems \cite{Distent}:
\begin{equation}
\tau_{3} = C_{A/BC} - (C_{A/B}+C_{A/C})
\label{t3pure}
\end{equation}

with $C_{A/BC}$ is  the I concurnce defined in (\ref{iconc}) and , $C_{A/B}$ and $C_{A/C}$ are the  Wooter's concurrence defined in  (\ref{Wconc})

\item For a general pure four qubits state:
\begin{equation}
\arrowvert  \psi^{ABCD}\rangle = \sum_{i_{1}i_{2}i_{3}i_{4}} a_{i_{1}i_{2}i_{3}i_{4}} \arrowvert i_{1}i_{2}i_{3}i_{4}\rangle
\label{psi4}
\end{equation}

The genuin tangle that quantify the 4-ways entanglement  for a pure state of four qubits is defined in \cite{12}:
\begin{equation}
\tau_{4} = 4 \vert [ (F_{0001}-F_{0000})  + (F_{0100}-F_{0011}) ] ^{2} \vert
\label{t4pure}
\end{equation}
in which   $F_{00i_{3}i_{4}} = det \begin{vmatrix}
a_{00i_{3}i_{4}} & a_{01i_{3+1}i_{4+1}} \\ 
a_{10i_{3}i_{4}} & a_{11i_{3+1}i_{4+1}}
\end{vmatrix}$.

\item For a symmetric pure four qubits states we can extract  the three tangle using the monogamy properties of entanglement \cite{9}:
\begin{equation}
	\tau_{3} = \dfrac{1}{3}(\tau_{1}  - \tau_{4} -3\tau_{2}),
\label{t3of4pure}
\end{equation}
with $\tau_{4},\tau_{2}$ and $\tau_{1}$  defined respectively in  (\ref{t4pure}), (\ref{to2}) and (\ref{to1}).

\item The lower Bound for mixed systems \cite{lb2009}  mentioned in (\ref{to1}, \ref{to3}, \ref{to4} and \ref{toN} ) is based in this version of concurrence :
\begin{equation}
	C_{\alpha\beta}^{p} = max \{ 0,\lambda(1)_{\alpha\beta}^{p}-\lambda(2)_{\alpha\beta}^{p}-\lambda(3)_{\alpha\beta}^{p}-\lambda(4)_{\alpha\beta}^{p}\}
	\label{clb}
\end{equation}
 In which $\lambda(i)_{\alpha\beta}^{p}$ being the square roots of the four nonzero eigenvalues, in decreasing order, of the non-Hermitian matrix $\sqrt{\rho\tilde{\rho}_{\alpha\beta}}$, with $\tilde{\rho}_{\alpha\beta} = (L_{\alpha} \otimes L_{\beta}) \rho^{*} (L_{\alpha} \otimes L_{\beta})$, and $L_{\alpha}$ and $L_{\beta}$ being the generators of $SO(d)$. \\

\end{itemize}

\section{Representative states in the Coherent states picture}
\label{appendix2}
Coherent state \cite{16} play an important role in quantum mechanics, specifically in quantum information resources as quantum computing and quantum circuit , because they are  easily realizable which in turn is due to the fact that they are the most classical quantum states of the harmonic oscillator, they are defined by: 
\begin{equation}
	\arrowvert \pm \alpha \rangle = e^{-\dfrac{\lvert \alpha \rvert^{2}}{2}} \sum_{n=0}^{\infty}\dfrac{(\pm \alpha)^{n}}{\sqrt{n!}} \arrowvert n \rangle 
\end{equation}
\\
In this work we use the coherent states basis and write the different perfect states of each classes, then we consider the following basic change:
\begin{equation}
	\arrowvert \pm \rangle = \dfrac{1}{\sqrt{2(1\pm e^{-2\lvert \alpha\rvert^{2}})}}(\arrowvert \alpha \rangle \pm \arrowvert - \alpha\rangle)
\end{equation}
Then:
\begin{equation}
\arrowvert GHZ_{3\alpha} \rangle = N_{3\alpha}^{GHZ}( \arrowvert \alpha , \alpha, \alpha  \rangle + \arrowvert -\alpha, -\alpha, -\alpha \rangle )
\label{GHZ3alpha}
\end{equation}
where $N_{3\alpha}^{GHZ}$ is the normalization factor:
\begin{equation}
N_{3\alpha}^{GHZ}=\dfrac{1}{\sqrt{2(\exp(-6\lvert \alpha\rvert^{2})+1)}}
\end{equation}
\begin{equation}
\arrowvert W_{3\alpha} \rangle = N_{3\alpha}^{W}( \arrowvert \alpha , \alpha, -\alpha  \rangle + \arrowvert \alpha , -\alpha, \alpha  \rangle \rangle + \arrowvert -\alpha , \alpha, \alpha \rangle)
\label{W3alpha}
\end{equation}
where $N_{3\alpha}^{W}$ is the normalization factor:
\begin{equation}
N_{3\alpha}^{W}= \dfrac{1}{\sqrt{3(2\exp(-4\lvert \alpha\rvert^{2})+1)}}
\end{equation}
\begin{equation}
\arrowvert GHZ_{4\alpha} \rangle = N_{4\alpha}^{GHZ} ( \arrowvert \alpha, \alpha , \alpha, \alpha  \rangle + \arrowvert -\alpha, -\alpha, -\alpha, -\alpha \rangle
\label{GHZ4alpha}
\end{equation}
where $N_{4\alpha}^{GHZ}$ is the normalization factor:
\begin{equation}
N_{4\alpha}^{GHZ}=\dfrac{1}{\sqrt{2(\exp(-8\lvert \alpha\rvert^{2})+1)}}
\end{equation}
\begin{equation}
	\arrowvert X_{4\alpha} \rangle = N_{4\alpha}^{X}( \arrowvert \alpha  , \alpha , \alpha, \alpha  \rangle + \arrowvert -\alpha , -\alpha, -\alpha  , \alpha \rangle + \arrowvert -\alpha , -\alpha, \alpha , -\alpha  \rangle + \arrowvert -\alpha , \alpha, -\alpha , -\alpha \rangle+ \arrowvert \alpha , -\alpha, -\alpha , -\alpha \rangle)
\label{X4alpha}
\end{equation} 
where $N_{4\alpha}^{X}$ is the normalization factor:
\begin{equation}
N_{4\alpha}^{X}=\dfrac{1}{\sqrt{5+8\exp(-6\lvert \alpha\rvert^{2})+12\exp(-4\lvert \alpha\rvert^{2}))}}
\end{equation}
\begin{equation}
\arrowvert W_{4\alpha} \rangle = N_{4\alpha}^{W}( \arrowvert -\alpha  , \alpha , \alpha, \alpha  \rangle + \arrowvert \alpha , -\alpha, \alpha  , \alpha \rangle + \arrowvert \alpha , \alpha, -\alpha , \alpha  \rangle + \arrowvert \alpha , \alpha, \alpha , -\alpha \rangle)
\label{W4alpha}
\end{equation}
where $ N_{4\alpha}^{W}$ is the normalization factor:
\begin{equation}
 N_{4\alpha}^{W}=\dfrac{1}{\sqrt{4(3\exp(-4\lvert \alpha\rvert^{2})+1)}}
\end{equation}

\end{appendices}

\end{document}